\documentclass[a4paper,11pt]{article}
\pdfoutput=1 % if your are submitting a pdflatex (i.e. if you have
\usepackage{jcappub} 
\usepackage[T1]{fontenc} % if needed
\usepackage{amsbsy}

\title{\boldmath Isocurvature modes: joint analysis of the CMB power spectrum and bispectrum}

\author[a,b]{T. Montandon,\note{Corresponding author.}}
\author[a]{G. Patanchon,}
\author[b]{B. van Tent}

\affiliation[a]{Universit\'e de Paris, CNRS, Astroparticule et Cosmologie, F-75006 Paris, France}
\affiliation[b]{Universit\'e Paris-Saclay, CNRS/IN2P3, IJCLab, 91405 Orsay, France}

\emailAdd{thomas.montandon@apc.in2p3.fr}
\emailAdd{guillaume.patanchon@apc.in2p3.fr}
\emailAdd{bartjan.van-tent@ijclab.in2p3.fr}

\abstract{We perform a joint analysis of the power spectrum and the bispectrum of the CMB temperature and polarization anisotropies to improve the constraints on isocurvature modes. We construct joint likelihoods, both for the existing Planck data, and to make forecasts for the future LiteBIRD and CMB-S4 experiments. We assume a general two-field inflation model with five free parameters, leading to one isocurvature mode (which can be CDM density, neutrino density or neutrino velocity) arbitrarily correlated with the adiabatic mode. We theoretically assess in which cases (of detecting and/or fixing parameters) improvements can be expected, to guide our subsequent numerical analyses. We find that for Planck, which detected neither isocurvature modes nor primordial non-Gaussianity, the joint analysis does not improve the constraints in the general case. However, if we fix additional parameters in the model, the improvements can be highly significant depending on the chosen parameter values. For LiteBIRD+CMB-S4 we study in which regions of parameter space compatible with the Planck results the joint analysis will improve the constraints or the significance of a detection. We find that, while for CDM isocurvature this region is very small, for the neutrino isocurvature modes it is much larger. In particular for neutrino velocity it can be about half of the Planck-allowed region, where the joint analysis reduces the isocurvature error bars by up to 70\%. In addition the joint analysis can also improve the error bars of some of the standard cosmological parameters, by up to 30\% for $\theta_{MC}$ for example, by breaking the degeneracies with the correlation parameter between adiabatic and isocurvature modes.}

\begin{document}
\maketitle
\flushbottom

\section{Introduction}
The Cosmic Microwave Background (CMB) radiation is a unique probe of the physics of the primordial universe. The small temperature and polarization fluctuations of the CMB result from fluctuations of the metric generated during inflation, an early phase of rapid exponential expansion of the universe. This means that the inflation paradigm can be tested by measuring the statistics of these CMB anisotropies. Inflation is driven, in the simplest models, by a single scalar field with a standard kinetic term. These models predict scalar perturbations that are adiabatic and nearly Gaussian, as well as tensor perturbations, which are both described by simple power-law spectra with a small tilt. These predictions for the scalar fluctuations are compatible with the Planck mission measurements \cite{Akrami:2018odb, Akrami:2019izv}. The tensor perturbations remain undetected today and their measurement is the main goal of future CMB experiments, like the LiteBIRD satellite mission \cite{Litebird,Hazumi:2019lys} and the CMB-S4 ground-based experiments \cite{Abazajian:2019eic,Abazajian:2016yjj}. Multi-field inflation, on the other hand, can generate a higher level of non-Gaussianity, see e.g.~\cite{Bartolo:2001cw,Bernardeau:2002jy,Lyth:2005fi,Rigopoulos:2005ae,Rigopoulos:2005us,Vernizzi:2006ve,Tzavara:2010ge,Jung:2016kfd} or the reviews \cite{Byrnes:2010em,Wang:2013zva}, that might be detectable with future CMB experiments as well as with future large-scale structure surveys, see e.g.~\cite{Karagiannis:2018jdt}. Multi-field inflation models are well-motivated in the context of high-energy theories of particle physics, given that these theories generally predict the presence of many scalar fields, see for example \cite{Lyth:1998xn}. Moreover, multi-field models can generate one or more isocurvature modes in addition to the adiabatic mode \cite{Bucher:1999re,Langlois:1999dw,Gordon:2000hv,GrootNibbelink:2001qt,vanTent:2003mn,Bassett:2005xm}. Even if the CMB is mostly adiabatic, isocurvature components can still be present at the level of $25\%$ at $2\sigma$ given the Planck measurements \cite{Akrami:2018odb}. For a very complete bibliography on mechanisms that can generate isocurvature modes, see \cite{Valiviita}.\par

Given that a detection of isocurvature modes would rule out single-field inflation as the sole source of the cosmological fluctuations, it is important to improve our constraints on these modes as much as possible. The Planck constraints mentioned above come from an analysis of the CMB power spectrum alone (the Fourier or spherical harmonic transform of the two-point correlation function). However, isocurvature modes can also have an impact on the CMB bispectrum \cite{Bartolo:2001cw,Kawasaki:2008sn,IsoNG1,IsoNG2,Hikage:2008sk,Langlois:2011zz,Langlois:2011hn,Kawakami:2012ke,Langlois:2012tm,Hikage:2012be,Hikage:2012tf} (the Fourier or spherical harmonic transform of the three-point correlation function). This was independently tested in the Planck bispectrum analysis \cite{Akrami:2019izv}, where no isocurvature non-Gaussianity was detected either. In this work, we perform a joint analysis of the power spectrum and the bispectrum to improve the constraints on isocurvature modes. Of course a joint analysis is only meaningful if the power spectrum and bispectrum observables are related, because they all depend on the same model parameters. To find a good compromise between generality on the one hand, and not adding too many new free parameters to our cosmology on the other hand, we assume a two-field inflation model, so that we have only a single isocurvature mode (of any type) in addition to the adiabatic mode. We also assume that one of the fields dominates both the linear isocurvature mode and the second-order (non-Gaussian) parts of the adiabatic and the isocurvature mode, the other field only contributing to the linear adiabatic mode. For the rest, however, this model is completely general. It is the same model as considered in the last section of \cite{Langlois:2012tm}.\par

For our joint analysis we construct a joint likelihood of the power spectrum and the bispectrum, both for Planck to analyze the existing data, and for LiteBIRD and CMB-S4 to make forecasts. We argue that this joint likelihood can be approximated as a simple multiplication of the individual likelihoods, as was also done in \cite{Meerburg:2015owa} for a joint analysis of resonant features in the power spectrum and bispectrum. For the bispectrum part we show that we can, without loss of performance, use a likelihood of the bispectrum amplitudes $\tilde{f}_\mathrm{NL}$.\par

This paper is organized as follows. In section~\ref{recap}, we briefly summarize how the adiabatic and isocurvature perturbation modes are defined and what types of isocurvature modes there are. In section~\ref{stats}, we first introduce the power spectrum and bispectrum observables independently. Then, we establish the link between them by introducing the two-field inflation model as discussed above. In section~\ref{jointa}, we discuss the power spectrum and bispectrum likelihoods for Planck and for LiteBIRD and CMB-S4, and how we combine them into a joint likelihood. In section~\ref{result}, we first show and discuss the results of our joint analysis of the Planck data, both for the general model and for the case where we fix additional parameters. We then give a general theoretical explanation of these results, that also provides guidance for the forecasts for future experiments. Finally we investigate in which Planck-allowed regions of the parameter space the joint analysis will improve the constraints on isocurvature modes (compared to an analysis of the power spectrum alone) for LiteBIRD+CMB-S4. We perform joint analyses for certain choices of fiducial parameters, also looking at the consequences for the other cosmological parameters. We summarize and conclude in section~\ref{concl}.

\section{Adiabatic and isocurvature modes}\label{recap}
Following \cite{Bucher:1999re,Bassett:2005xm,Langlois:1999dw}, we define in this section the adiabatic mode and the isocurvature modes of the scalar perturbations. In general for each matter component of the universe (photons, baryons, cold dark matter (CDM), and neutrinos) there will be both an energy/matter density fluctuation and a velocity fluctuation, both sourced by the fluctuations of the scalar inflationary fields from which these components originate. The density contrast of component $i$, $\delta_i(t,\mathbf{x})$, is defined as the density fluctuation $\delta \rho_i(t,\mathbf{x})$ divided by the mean density $\rho_i(t)$. The velocity fluctuation, in the case of scalar fluctuations, will only have a curl-free component that is described by its divergence: $v_i(t,\mathbf{x}) = \mathbf {\nabla}  \cdot \mathbf{v}_i$. During the radiation era that follows the reheating era, one can write the evolution equations for $\delta$ and $v$. The solutions of these equations need initial conditions, which are the modes generated by inflation. We also define the total curvature perturbation first defined in \cite{Bardeen:1983qw} (see also \cite{martinschwarz,Wands:2000dp}), often called $\zeta$, which is a gauge-invariant quantity:
 \begin{equation}
 \label{zeta}
     \zeta = -\psi -H \frac{\delta\rho}{\dot \rho}
 \end{equation}
where $\psi$ is defined by writing that part of the spatial metric $g_{ij}$ that is proportional to $\delta_{ij}$ as $a^2(1-2\psi)\delta_{ij}$ up to first order, with $a$ the scale factor; $H=\dot{a}/a$ is the Hubble parameter and the dot stands for a time derivative. We talk about ``curvature'' since the Laplacian of $\psi$ is proportional to the spatial curvature (and so is the Laplacean of $\zeta$ itself in a gauge where $\delta\rho=0$).\par
Usually, we decompose the initial condition into two different perturbations: curvature and isocurvature initial perturbations. By initial condition we mean deep in the radiation era on super-horizon scales.
Actually, imposing adiabatic initial conditions is equivalent to $\zeta^0 \neq 0$, i.e.\ a curvature initial perturbation. Adiabatic initial conditions imply that the number density perturbations of all components are fluctuating in phase: $\delta n_i/n_i = \delta n_j/n_j$. This is equivalent to the following relations for the initial density contrasts:
\begin{equation}
\label{adiabatique}
\delta_c^0 = \delta_b^0 = \frac{3}{4}\delta_{\nu}^0  = \frac{3}{4}\delta_{\gamma}^0
\end{equation}
where the $^0$ stands for initial condition and $c,b,\nu$ and $\gamma$ refer to cold dark matter, baryon, neutrino and photon, respectively. In the adiabatic initial conditions, all species have zero velocities.
\par
On the other hand, we can also have isocurvature initial perturbations, where the total curvature perturbation initially vanishes: $\zeta^0 = 0$. There are two possibilities to have isocurvature modes. The first is by perturbing the energy density of a species compared to another such that the total fluctuation remains zero. Usually, we break one of the equalities in \eqref{adiabatique} and we always take the photon density as reference. It means:
\begin{equation}
\label{density}
S_i^d = \frac{1}{1+\omega_i} \delta_{i \neq \gamma}^0 - \frac{3}{4}\delta_{\gamma}^0
\end{equation}
where we have introduced $\omega_i$ which is the ratio of the pressure divided by the energy density of the particular species $i$. The factor $3/4$ is the result of $1/(1+\omega_{i})$ for the photon where $\omega_{i=\gamma}=1/3$. Because this breaks the adiabaticity, we often call $S_i^d$ the entropic perturbation, which is a gauge-invariant quantity. If we have $N$ scalar fields during inflation, then we can have at most $N-1$ entropic perturbations and 1 adiabatic mode. 
Finally, as shown in detail in \cite{CDM-BAR}, the CDM and baryon isocurvature modes are indistinguishable in the CMB power spectrum (they only differ in amplitude). Therefore, we will not consider the baryon isocurvature mode in this paper. 
\par The second category of isocurvature perturbations is the velocity type. It corresponds to a universe with no initial density perturbations $(\delta_i = 0)$, but the species do not have the same initial velocities. We define 
\begin{equation}
\label{velocity}
S_{i}^v = \frac{1}{1-f_{\nu}}(v_{i}^0  -v_{\gamma}^0)
\end{equation}
where $f_{\nu}$ is the fraction of neutrino density with respect to the total radiation density. There are also $N-1$ velocity modes. Let us first consider the baryon velocity initial condition. Those perturbations may exist initially but the tight coupling of the baryons with the photons will make $S_{b}^v \rightarrow 0$ very rapidly given the considered time scales. It is then useful to always consider $v_b = v_{\gamma}=v_{\gamma b }$. Next, the CDM velocity mode is always zero. We can see that by switching to the synchronous gauge. The synchronous gauge is defined by free-falling observers, so it is always possible to use the CDM particles to define the coordinates, since these particles interact only through gravity. Thus, in this gauge $v_c=0$, see \cite{Ma:1995ey}. Furthermore, we can show that imposing $\delta_i = 0$ in this gauge fixes the baryon/photon velocity to zero. It follows that in all gauges $S_{c}^v=0$, as it is gauge-invariant. In the end, the only velocity isocurvature mode that we have to consider is the neutrino one. \par

To conclude, we have the adiabatic mode and 3 isocurvature modes to consider: the CDM density mode, the neutrino density mode, and the neutrino velocity mode. This decomposition of the initial modes coming from inflation is useful, because it helps us to differentiate single-field from multi-field inflation. Single-field inflation can only generate an adiabatic mode while multi-field models can generate adiabatic and isocurvature initial perturbations. To test this, we will look at the CMB temperature and polarization fluctuations. In practice we will always consider only a single isocurvature mode in addition to the adiabatic mode (which can be correlated), because otherwise the number of free parameters, in particular for the bispectrum, would be too high to get meaningful constraints.

\section{Two- and three-point statistics}\label{stats}
To study the different perturbations generated by inflation, which we call primordial perturbations, we will look at the CMB anisotropies through their 2-point and 3-point statistics. At the end of inflation, the distribution of fields is usually supposed to be Gaussian. Nearly Gaussian fields are well characterized by their power spectra:
\begin{equation}
\label{parimordialsp}
    (2\pi)^3 \delta(\mathbf{ k_1} +\mathbf {k_2}) P^{IJ}(k_1) = \left< I(\mathbf{ k_1}) J(\mathbf{ k_2}) \right>
\end{equation}
where $I$, $J$ and $K$ stand for curvature $(\zeta)$ or any possible isocurvature modes $(S)$. The vector $\mathbf {k}$ is the 3-dimensional momentum. Deviations from Gaussianity can be studied via the primordial bispectra:
\begin{equation}
\label{parimordialbi}
    (2\pi)^3 \delta(\mathbf{ k_1} +\mathbf {k_2}+\mathbf {k_3}) B^{IJK}(k_1,k_2,k_3) = \left< I(\mathbf{ k_1}) J(\mathbf{ k_2}) K(\mathbf{ k_3}) \right>
\end{equation}
In this section, we will briefly describe the standard way of studying these quantities via the observations of temperature and polarization of the CMB.

\subsection{Power spectrum}
In \cite{Akrami:2018odb,Ade:2015lrj}, the primordial power spectra are modeled as power laws with the amplitudes fixed at two pivot scales $k_1$ and $k_2$ (we use this parametrization instead of the more usual choice of one amplitude and a spectral index):
\begin{equation} \label{primordial}
    \mathcal P^{IJ}(k) = \exp \left( \frac {\ln(k)-\ln(k_2)}{\ln(k_1)-\ln(k_2)} \ln\left( \mathcal  P^{(1)}_{IJ}\right) + \frac {\ln(k)-\ln(k_1)}{\ln(k_2)-\ln(k_1)} \ln\left( \mathcal P^{(2)}_{IJ}\right) \right)
\end{equation}
where $I$ and $J \in [\zeta, S]$ (we consider only a single isocurvature mode in addition to the adiabatic one) and $\mathcal P^{(1)}_{IJ}$ and $\mathcal P^{(2)}_{IJ}$ are the amplitudes at the two pivot scales. The two pivot scales are chosen to cover most of the observable range of Planck: $k_1=0.002$~Mpc$^{-1}$ and $k_2= 0.1$~Mpc$^{-1}$. The choice of the parametrization is related to the question of priors. Indeed, a change of parametrization, which means changing the priors, gives a different posterior distribution when maximizing the likelihood. Applying a flat prior on the usual amplitudes and spectral indices is not equivalent to applying a flat prior on $\mathcal P^{(1)}_{IJ}$ and $\mathcal P^{(2)}_{IJ}$. The usual approach with amplitudes and spectral indices as free parameters produces a strongly prior-dependent posterior for the isocurvature modes especially if the isocurvature spectral index is free, given the absence of a significant detection, and is then difficult to interpret as explained in \cite{Planck:2013jfk}. Unless specified differently, we will always apply flat priors on $\mathcal P^{(1)}_{\zeta \zeta}$, $\mathcal P^{(2)}_{\zeta \zeta}$, $\mathcal P^{(1)}_{\zeta S}$ and $\mathcal P^{(1)}_{SS}$. \par
Since the bispectrum parametrization with a free isocurvature spectral index would have too many parameters to be constrained in the near future, we study the case where $n_{s,iso} = n_{s,adi} = n_{s,cross} =n_s$, which is for example motivated by the curvaton scenario. This restriction imposes:
\begin{equation}\label{ns}
\mathcal P_{SS}^{(2)} = \frac{\mathcal P_{\zeta \zeta}^{(2)}}{\mathcal P_{\zeta \zeta}^{(1)}} \mathcal P_{SS}^{(1)},
\qquad \qquad
\mathcal P_{\zeta S}^{(2)} = \frac{\mathcal P_{\zeta \zeta}^{(2)}}{\mathcal P_{\zeta \zeta}^{(1)}} \mathcal P_{\zeta S}^{(1)}
\end{equation}
It is convenient to define two parameters also used in \cite{Akrami:2018odb, Ade:2015lrj}:
\begin{equation} \label{param_planck}
     \beta_{\rm iso} = \frac{\mathcal P_{SS}^{(i)}}{\mathcal P^{(i)}_{\zeta\zeta}+\mathcal P^{(i)}_{SS}}, 
     \qquad \qquad 
     \cos{\Delta} = \frac{\mathcal P^{(i)}_{\zeta S}}{\sqrt{\mathcal P^{(i)}_{\zeta \zeta}\mathcal P^{(i)}_{S S}}}
\end{equation}
Here $\beta_{\rm iso}$ is the ratio of the isocurvature amplitude to the sum of all amplitudes and $\cos \Delta$ is the relative correlation of the adiabatic mode and the isocurvature mode. The parameters $\beta_{iso}$ and $\cos \Delta$ do not have an $(i)$ index since they are independent of the pivot scale. The prior of $\mathcal P^{(1)}_{\zeta S}$ has to be chosen such that the second equation of \eqref{param_planck} can be satisfied, i.e.\ $(\mathcal P^{(i)}_{\zeta S})^2 \leq \mathcal P^{(i)}_{\zeta \zeta}\mathcal P^{(i)}_{S S} $. Thanks to our hypothesis on the spectral indices, the isocurvature parameters defined in \eqref{param_planck} are independent of the pivot scale (hence $(i)$ can be either $(1)$ or $(2)$). 

Once we have the primordial power spectra in \eqref{parimordialsp}, we have to take into account their evolution through the radiation era until the emission of the CMB using the Einstein equations and translate them into observable CMB quantities using the Boltzmann equations. This step is done using existing Boltzmann codes that we will specify later. This leads to the CMB power spectra:
\begin{equation}
\label{Cliso}
C^{\lambda_1\lambda_2,IJ}_{\ell}=\left<  a_{\ell m}^{\lambda_1,I} a_{\ell m}^{\lambda_2,J} \right> = \frac{2}{\pi}\int_0^\infty dk \, k^2 g_{\ell}^{\lambda_1,I}(k) g_{\ell}^{\lambda_2,J}(k) P^{IJ}(k)
\end{equation}
where  $g_{\ell}^{\lambda_1,I}(k)$ are the radiation transfer functions, and $\lambda_1$ and $\lambda_2$ are indices for temperature or polarization of the CMB. The theoretical total angular power spectrum of the CMB is:
\begin{equation}
\label{Cltot}
     C_{\ell}^{th,\lambda_1\lambda_2} =  A_s^{\zeta \zeta} \left(  \bar C_{\ell}^{\lambda_1\lambda_2,\zeta \zeta} + \alpha \, \bar C_{\ell}^{\lambda_1\lambda_2,SS} + 2 \cos{\Delta} \sqrt{\alpha} \,  \bar C_{\ell}^{\lambda_1\lambda_2,\zeta S} \right)
\end{equation}
where $\bar C_{\ell}$ are the  normalized power spectra with $A_{s}^{IJ}=1$. For simplicity we have used here the usual parameters $A_s^{IJ}$ which are the amplitudes of the primordial power spectra at $k_0=0.05$~Mpc$^{-1}$. We have also introduced $\alpha \equiv \beta_{\rm iso}/(1-\beta_{\rm iso})$. We will always use $\alpha$ in the theoretical part because expressions are more compact. But in the analyses, we will use $\beta_{\rm iso}$ so that we can compare our results with Planck.

\subsection{Bispectrum}
Following \cite{Langlois:2012tm}, we will assume the local shape for the primordial bispectrum. This choice is theoretically justified by the fact that multi-field inflation, which are the models that generate isocurvature modes, also generate non-Gaussianities of the local shape. Here, we consider one isocurvature mode in addition to the adiabatic mode (otherwise there would be too many free parameters and no meaningful constraints could be obtained). We then use a generalized form of the local shape. Again, we assume all spectral indices to be equal. The primordial bispectra, for each tuple of $IJK$, can be expressed as a sum of terms quadratic in the adiabatic power spectrum (see \cite{Langlois:2012tm}):
\begin{equation}
B^{IJK}(k_1,k_2,k_3) = \tilde f_{\rm NL}^{I,JK} P_{\zeta}(k_2) P_{\zeta}(k_3)+\tilde f_{\rm NL}^{J,KI} P_{\zeta}(k_1) P_{\zeta}(k_3)+\tilde f_{\rm NL}^{K,IJ} P_{\zeta}(k_1) P_{\zeta}(k_2)
\end{equation}
The comma in $(I,JK)$ indicates that the order of the last two indices is not important because of a local bispectrum symmetry (invariance under simultaneous interchange of e.g.\ $J$ and $K$ and $k_2$ and $k_3$). We use $\tilde{f}_{\rm NL}$ (with a tilde) as they are defined in terms of $\zeta$, as in \cite{Langlois:2012tm}, to avoid confusion with the more usual $f_{\rm NL}$, used e.g.\ in the Planck results \cite{Ade:2013ydc,Ade:2015ava}, which are defined in terms of the gravitational potential $\psi$.  

Again, the evolution of the primordial bispectrum from the radiation era until the CMB emission is described by the Einstein-Boltzmann equations. For the CMB, instead of looking at the full bispectrum $B_{\ell_1\ell_2\ell_2}^{m_1 m_2 m_3}$ defined equivalently to the first equality in \eqref{Cliso} but with three different $a_{\ell m }$, we can factorize the $m$ dependence into a Gaunt integral and then only look at the reduced bispectrum  $b_{\ell_1\ell_2\ell_2} = B_{\ell_1\ell_2\ell_2}^{m_1 m_2 m_3}/ G_{\ell_1\ell_2\ell_2}^{m_1 m_2 m_3}$. For the CMB, it takes the form:
\begin{equation}
\begin{split}
b^{\lambda_1\lambda_2\lambda_3;IJK}_{\ell_1\ell_2\ell_3} =\left(\frac{2}{\pi}\right)^3 \int \left(\prod_{i=1}^3 dk_i \, k_i^2 \right) &g_{\ell_1}^{I;\lambda_1}(k_1) g_{\ell_2}^{J;\lambda_2}(k_2) g_{\ell_3}^{K;\lambda_3}(k_3) B^{IJK} (k_1,k_2,k_3)\\
&\times \int_0^\infty r^2dr j_{\ell_1 }(k_1r) j_{\ell_2 }(k_2r) j_{\ell_3 }(k_3r)
\end{split}
\end{equation}
where the functions $j_{\ell }(kr)$ are spherical Bessel functions due to the projection of the three-dimensional Fourier modes onto a two-dimensional sphere. We can use the expression of the total reduced bispectrum given in \cite{Langlois:2012tm}:
\begin{equation}
\label{bisp}
b_{\ell_1\ell_2\ell_3} = \sum_{IJK} \tilde f_{\rm NL}^{I,JK}b^{I,JK}_{\ell_1\ell_2\ell_3}
\end{equation}
We see in this formula how the $\tilde f_{\rm NL}$ factorize and how they can be interpreted as the amplitudes of each normalized bispectrum.  There are six different $\tilde f_{\rm NL}$ (and not eight) because of the symmetry of the local shape. The total angular bispectrum  is a function of these six $\tilde f_{\rm NL}$ parameters and of the cosmological parameters through the transfer functions: the current baryon and CDM densities $\Omega_b, \Omega_c$, the sound horizon at recombination $\theta_{MC}$ and the re-ionisation optical depth $\tau$. In the rest of the paper we will denote these four cosmological parameters together with $A_s^{\zeta\zeta}$ and $n_s$ as $\boldsymbol{\theta}$.

\subsection{Two-field model: link between parameters}
\label{two-field}
To perform a joint analysis, we assume a model with two scalar fields acting during inflation: $\phi$ and $\sigma$. This model is crucial for our analysis, because it provides the link between what is measured in the bispectrum and in the power spectrum. Without such a theoretical link there would be no point in a joint analysis. We want the model to be as general as possible, so that our analysis applies to as broad a class of models as possible, while at the same time we must restrict the number of additional free parameters to just a few in order to get any meaningful constraints. In the most general non-Gaussian two-field model, up to second order in cosmological perturbation theory, we would have 10 parameters to constrain. This would be too much to obtain meaningful results given the current experimental constraints. To reduce the number of parameters, we make the following two assumptions (similar to the model in the final section of \cite{Langlois:2012tm}): we suppose that the isocurvature mode is dominated by the contribution of one field, which we assume to be $\phi$, and that this same field $\phi$ also dominates the second-order terms for both the adiabatic and the isocurvature mode. The field $\sigma$ then only contributes to the linear adiabatic mode. These two assumptions remove 5 parameters and allow for a good compromise between generality and number of free parameters.
We can then write, up to second order:
\begin{equation}\label{ourmodel}
   \zeta = \zeta_{\sigma} \delta \sigma + \zeta_{\phi} \delta \phi + \frac{1}{2} \zeta_{\phi\phi} \delta \phi ^2, \quad \quad \quad \quad
    S =  S_{\phi} \delta \phi + \frac{1}{2} S_{\phi\phi} \delta \phi ^2
\end{equation}

With the usual inflationary assumptions, the field perturbations $\delta \phi$ and $\delta \sigma$ can be considered independent and quasi-Gaussian with the same power spectrum. We can then calculate $\left< I(\mathbf k_1)J(\mathbf k_2) \right>$ where $I,J \in [\zeta,S]$. Using \eqref{Cltot} and \eqref{parimordialsp}, we can establish the link between the parameters of the model and the analysis parametrization of \eqref{param_planck}:
\begin{equation}
    \label{alpha}
    A_s^{\zeta \zeta} = \zeta_{\phi}^2 + \zeta_{\sigma}^2, \qquad  \alpha = \frac{\beta_{\rm iso}}{1-\beta_{\rm iso}} = \frac{S_{\phi}^2}{\zeta_{\phi}^2+\zeta_{\sigma}^2}, \qquad  \cos{\Delta} = \frac{\zeta_{\phi}}{\sqrt{\zeta_{\phi}^2+\zeta_{\sigma}^2}}
\end{equation}
Calculating the six correlations $\left< I(\mathbf k_1)J(\mathbf k_2) K(\mathbf k_3) \right>$, we can express the $\tilde f_{\rm NL}$ as follows:
\begin{equation}
  \label{fnl}
  \begin{aligned}
    \tilde f^{\zeta,\zeta\zeta}_{\rm NL} &= \kappa_{\zeta} \cos^2\Delta  \\
    \tilde f^{\zeta,\zeta S} _{\rm NL}   &= \kappa_{\zeta} \cos \Delta \sqrt{\alpha} \\
    \tilde f^{\zeta,SS}_{\rm NL} &= \kappa_{\zeta} \,\alpha
  \end{aligned}
  \qquad \qquad
  \begin{aligned}
    \tilde f^{S,\zeta\zeta}_{\rm NL} &= \kappa_{S} \cos^2\Delta  \\
    \tilde f^{S,\zeta S} _{\rm NL}   &= \kappa_{S}\cos \Delta \sqrt{\alpha} \\
    \tilde f^{S,SS}_{\rm NL} &= \kappa_{S} \,\alpha
  \end{aligned}
\end{equation}
where we have defined the two $\kappa_I$ as the coefficients of the second-order terms normalized by the adiabatic power spectrum:
\begin{equation}\label{defkappa}
    \kappa_{\zeta} = \frac{\zeta_{\phi\phi}}{\zeta_{\sigma}^2+\zeta_{\phi}^2}, \qquad \qquad
  \kappa_{S} = \frac{S_{\phi\phi}}{\zeta_{\sigma}^2+\zeta_{\phi}^2}
\end{equation}
This parametrization differs from \cite{Langlois:2012tm}. Our parametrization has the advantage that it remains valid in the limit of uncorrelated modes, $\cos \Delta \rightarrow 0$.\footnote{In \cite{Langlois:2012tm}, they use $\mu_I$ instead of $\kappa_I$ where $\mu_I  = \kappa_I/\cos^2 \Delta$. We find immediately that $\mu_I$ is infinite when $\cos \Delta=0$. Furthermore, we inverted the symbols used for the fields so that in the single-field limit, the single field is $\phi$ instead of $\sigma$. Finally, they use a parameter $\Xi$ which is related to $\cos\Delta$ as $\cos \Delta = \epsilon_{\zeta S} \sqrt{\Xi}$ with $\epsilon_{\zeta S}=\pm1$ denoting the relative sign of $\zeta$ and $S$.} 
Looking at these equations \eqref{fnl}, we see that the $\tilde f^{I,\zeta S} _{\rm NL}$ are proportional to the amplitude of the cross-power spectrum $C^{\zeta S}_{\ell}$ and that the $\tilde f^{I,S S} _{\rm NL}$ are proportional to the amplitude of the pure isocurvature power spectrum $C^{S S}_{\ell}$.
In this model, the single-field limit corresponds to $\delta \sigma \rightarrow 0$. Of course, in this limit we must have a zero isocurvature component, i.e.\  $S_{\phi} = S_{\phi \phi}=0$. All of this translates into $\cos \Delta = 1$ and $\alpha = \kappa_S = 0$. In equations \eqref{fnl}, it means that the only non-zero quantity is $\tilde f_{\rm NL}^{\zeta,\zeta\zeta}$ in that case.  \par
The power spectrum \eqref{Cltot} depends on the 6 $\Lambda$CDM parameters.  Two of these parameters are related to the inflationary adiabatic mode, $A_s$ and $n_s$, which are in this paper derived parameters from $\mathcal P_{\zeta\zeta}^{(1)}$ and $\mathcal P_{\zeta\zeta}^{(2)}$. We also have two additional parameters related to the isocurvature mode, the relative amplitude $\alpha$ and the correlation parameter $\cos \Delta$ derived from $\mathcal P_{SS}^{(1)}$, $\mathcal P_{\zeta S}^{(1)}$ (and $\mathcal P_{\zeta\zeta}^{(1)}$). In the following we will call $\mathbf{\tilde f_{\rm NL}}$ the vector of the 6 $\tilde f_{\rm NL}$.
Using this model, the bispectrum, which is a function of 12 parameters (the 6 parameters $\boldsymbol{\theta}$ and the 6 parameters $ \mathbf{\tilde f_{\rm NL}}$), can be reduced to a function of 10 parameters (the 6 parameters $\boldsymbol{\theta}$ plus $\kappa_\zeta, \kappa_S, \alpha, \cos\Delta$). Only three of the $\tilde f_{\rm NL}$ are independent since we can easily find these three relations:
\begin{equation}\label{fnllink}
\begin{aligned}
\tilde f^{I,\zeta\zeta}_{\rm NL} \tilde f^{I,SS}_{\rm NL} = \left( \tilde  f^{I,S\zeta}_{\rm NL}\right)^2,
\end{aligned}
\qquad \qquad
\begin{aligned}
\tilde f^{\zeta,\zeta\zeta}_{\rm NL} \tilde f^{S,SS}_{\rm NL} = \tilde f^{S,\zeta\zeta}_{\rm NL} \tilde f^{\zeta,SS}_{\rm NL}
\end{aligned}
\end{equation}
where the first equation contains two relations for $I \in [\zeta,S]$. At the end, we have three independent $\tilde f_{\rm NL}$ which are expressed in terms of the four parameters $\{\kappa_{\zeta}, \kappa_S, \alpha, \cos \Delta\}$. Hence, the system is under-determined. 
The first equation in \eqref{fnllink} also offers a simple refutable prediction of the model: $\tilde f^{I,\zeta\zeta}_{\rm NL}$ and $\tilde f^{I,SS}_{\rm NL}$ must share the same sign, i.e.\ the sign of $\kappa_I$. The second equation does not add more information on the signs. Looking at equations \eqref{fnl}, we also have constraints on the sign of the correlation: if $\kappa_I>0$ ($\kappa_I<0$) then $\tilde f^{I,\zeta S}_{\rm NL}$ must have the same (opposite) sign as the correlation $\cos\Delta$.

\section{Joint analysis methodology} \label{jointa}

In this section, we describe the combination of the power spectrum likelihood with a ``bispectrum likelihood'' to perform the joint analysis, as well as its implementation. 

\subsection{Joint likelihood}

Rigorously, the power spectrum and bispectrum estimators are not statistically independent since they are calculated from the combination of the same modes in the observed maps. However, the calculation of the two estimators involves the linear combination of a large number of pairs and triplets of $a_{\ell m}$ modes (they are averaged over all multipole moments $m$ and in large bins of multipoles $\ell$) leading to nearly Gaussian statistics of the estimated power spectra and bispectra in the limit of weak non-Gaussianity of the CMB. Consequently, the cross-correlation of the two- and three-point functions, which involves averaging a large number of products of five Gaussian $a_{\ell m}$, vanishes, so that the estimators are uncorrelated. The independence of the two estimators can also be assumed since higher-order statistics are negligible because of the nearly Gaussian statistics of the power spectra and bispectra. We can then multiply the two distributions to obtain the total likelihood. The independence of the two-point and the three-point statistics has also been stressed in \cite{Meerburg:2015owa,Fergusson:2014tza,Fergusson:2014hya}. The power spectrum likelihood $\mathcal{L}$ is a function of all cosmological parameters stored in  $\boldsymbol{\theta}$ and of ($\alpha, \cos \Delta$). The bispectrum likelihood $P$ is a function of  $\mathbf{\tilde f_{\rm NL}}$. Thanks to \eqref{fnl}, we translate the $\tilde f_{\rm NL}$ into $\boldsymbol{\xi}=(\alpha,\cos\Delta,\kappa_{\zeta},\kappa_{S} )$ such that
\begin{equation}\label{joint}
    \mathbf{\mathcal{L}}^{tot} (\boldsymbol{\theta}, \boldsymbol{\xi}) =\mathbf{\mathcal{L}}(\boldsymbol{\theta}, \alpha, \cos \Delta) \times P(\boldsymbol{\xi})
\end{equation}
Both likelihoods will be further specified in the next sections.

\subsection{Power spectrum likelihood}

Let us first consider the power spectrum likelihood for different experiments: Planck, LiteBIRD, and CMB-S4.

\subsubsection{Planck}\label{Planck likelihood}
We use the Planck likelihood 2018 described in  \cite{Aghanim:2019ame}. This likelihood combines two parts covering two different multipole $\ell$ ranges. The low-$\ell$ part contains the multipoles lower than 30, and a pixel-based likelihood is used to account for non-stationarity of the signal and noise. This likelihood assumes Gaussian statistics for the maps.
At high $\ell$ the accurate calculation of the pixel covariance matrix is impossible. The likelihood for $\ell\geq 30$ can be approximated assuming Gaussian statistics of the power spectra, since each of the power spectrum amplitudes is estimated with a large number of modes.
This is an approximation of the likelihood, in particular because the observations are not full sky due to the galactic cut to mask foregrounds. Thus the Planck likelihood uses pseudo-$C_{\ell}$'s which are calculated on a masked sky. On the rest of the sky, which is used for the analysis, foreground parameters are estimated using a multi-component model of the power spectra. The total number of estimated parameters, taking into account all nuisance parameters, is 27. \par
We also include the lensing term as in the Planck 2018 likelihood, see \cite{Aghanim:2018oex}. They used the quadratic estimator developed by \cite{Okamoto:2003zw} to reconstruct the lensing field from the statistical anisotropies of the temperature and polarization fields. They then estimated the lensing power spectrum from the connected part of the CMB trispectrum. The lensing likelihood is supposed to be Gaussian with respect to the power spectrum.  \par
Currently, using the Planck power spectrum only, the $2 \sigma$ upper limits for $\beta_{\rm iso}$ are $0.039$, $0.089$, and $0.058$, for CDM, neutrino density and neutrino velocity, respectively. For $\cos \Delta$ the $2\sigma$ intervals are $[-0.41,0.31]$, $[-0.18,0.19]$ and $[-0.25,0.06]$, respectively. 

\subsubsection{LiteBIRD}

\begin{table}[t]
\centering
\begin{tabular}{|c |c c| c| }
 \hline
 \multicolumn{4}{ |c| }{LiteBIRD} \\
 \hline
 Channel [GHz] & \multicolumn{2}{ |c| }{ Noise [$\mu$K$^2$]} & Beam FWHM [arcmin] \\
 &Temperature  & Polarization&  \\
  \hline
119&3.58e-06 &1.43e-05&23.7\\
140&2.29e-06 &9.17e-06&20.7\\
100&5.45e-06 &2.18e-05&37.0\\
119&3.58e-06 &1.43e-05&31.6\\
140&2.29e-06 &9.17e-06&27.6\\
166&4.60e-06 &1.84e-05&24.2\\
195&3.12e-06 &1.24e-05&21.7\\
235&5.75e-06 &2.30e-05&19.6\\
 \hline
\end{tabular}
\caption{\label{litebird} We summarize in this table the characteristics of the LiteBIRD experiment. The values are taken from \cite{Litebird}. In the left column, the frequency channels where the CMB emission is dominant. In the second column, the amplitude of the noise power spectrum for temperature and polarization, and in the third column the beam size of the instrument.}
\end{table}

After the analysis of the Planck data, we will study the forecasts for future experiments. LiteBIRD is the next spatial mission for the observation of the CMB \cite{Hazumi:2019lys}, and has been selected by JAXA as a Strategic Large-Class mission to be launched in 2027. The first step is to compute the expectation of the observed power spectrum. To do so, let us define the matrix of the fiducial power spectra $\mathbf {C_{\ell}^{fid}}$, i.e. the theoretical power spectra calculated for the fiducial or 'true' parameters, that can be calculated through:
\begin{equation}
\label{Clobs}
C^{\lambda_1\lambda_2,\mathrm{fid}}_{\ell}=\left<  s_{\ell m}^{\lambda_1,\mathrm{fid}} s_{\ell m}^{\lambda_2,\mathrm{fid}} \right>
\end{equation}
where $s_{\ell m}^{\lambda_2,\mathrm{fid}}$ is the fiducial pure signal without noise (including only CMB and no foreground in our analysis), and $\left<.\right>$ is the expectation. The diagonal is composed of the auto-correlation $TT$, $EE$ and the lensing $\Phi\Phi$ power spectra and the non-zero off-diagonal terms are the cross-correlations $TE = ET$. We neglect here the correlation spectra $\Phi T$ and $\Phi E$. In \cite{DiValentino:2016foa} it was shown that considering unlensed spectra does not change the result for CORE significantly, which was a more sensitive experiment which has not been selected by ESA. For more details about CORE, see \cite{Delabrouille:2017rct}. Given these results we only consider lensed power spectra.\par
By adding the noise and the beam, we define the matrices $\mathbf { \tilde C_{\ell}^{X}}$ as follows:
\begin{equation}
\label{clobs}
\tilde C_{\ell}^{X,\lambda_1\lambda_2} = C_{\ell}^{X,\lambda_1\lambda_2} +  \left(h_{\ell}^{\lambda_1\lambda_2}\right)^{-2} n_{\ell}^{\lambda_1\lambda_2}
\end{equation}
where $X$ means either fiducial or theoretical, the last being calculated for any parameter set (and not only fiducial) and defined in equation~\eqref{Cltot}, $n_{\ell}^{\lambda_1\lambda_2}$ is the $\lambda_1\lambda_2$-component of the noise matrix, which is diagonal because we assume that there are no correlations between the temperature and the polarization noise, and $h_{\ell}^{\lambda_1\lambda_2}$ is the beam transfer function.\footnote{Usually, the beam is denoted by $b_{\ell}$. But to avoid any possible confusion with the reduced bispectrum in \eqref{bisp}, we choose to change this notation to $h_{\ell}$.} We use the noise specifications for LiteBIRD from \cite{Litebird} given in table \ref{litebird}. We assume that the channels are combined by weighting with the inverse noise variance, neglecting the effect of component separation. We assume that each channel has white noise, $n_{\ell}^{\lambda_1\lambda_2,f}$, and a Gaussian beam, $h_{\ell}^{\lambda_1\lambda_2,f}$. We combine them to obtain the second effective term of equation \eqref{clobs} as follows:
\begin{equation}
\label{exp}
\left(h_{\ell}^{\lambda_1\lambda_2}\right)^{-2} n_{\ell}^{\lambda_1\lambda_2} = \left( \sum_f  \frac{\left( h_{\ell}^{\lambda_1\lambda_2,f}\right)^2}{n_{\ell}^{\lambda_1\lambda_2,f}} \right)^{-1}
\end{equation}

The beam-convolved noise of the lensing power spectrum has a different origin, because we measure it indirectly using temperature and polarization. In \cite{Challinor:2017eqy}, approximations of the lensing noise at large scales are given for the temperature and for the polarization estimators. Given these approximations, the most powerful estimator for LiteBIRD is the one obtained from $E$ and $B$ correlations. For simplicity and because it is the most powerful estimator, we only consider the $EB$ estimator. The approximation on large scales of the noise given in \cite{Challinor:2017eqy} is:
\begin{equation}
\label{NP}
\left(h_{\ell}^{\Phi \Phi}\right)^{-2} n_{\ell}^{\Phi \Phi} = \left( \frac{\ell^4}{2} \sum_{\ell} \frac{2\ell+1}{4\pi} \frac{ \left(C_{\ell}^{\mathrm{fid},EE}\right)^2 }{\tilde C_{\ell}^{\mathrm{fid},EE} \tilde C_{\ell}^{\mathrm{fid},BB}}\right)^{-1} 
\end{equation}
In \cite{Challinor:2017eqy}, it is shown that for CORE this approximation is valid for low $\ell$. We place a cutoff for LiteBIRD at $\ell = 165$. This value is obtained by simply multiplying $\ell=550$, which was found in \cite{Challinor:2017eqy} for CORE, with the ratio of the beam FWHM for the two experiments. 

The fiducial power spectra are created after making a choice for the cosmological parameters $\boldsymbol{\theta}$ as well as for the isocurvature parameters, by using \eqref{clobs}, as will be discussed later. We can then fit our theoretical power spectrum defined in \eqref{Cltot} to the fiducial one using the full-sky likelihood as in \cite{DiValentino:2016foa}:
\begin{equation} \label{corelik}
\mathcal{L}(\boldsymbol{\theta},\alpha,\cos{\Delta}) = - \frac{1}{2} \sum_{\ell} f_\mathrm{sky} (2\ell+1) \left[ \frac{D_{\ell}}{|\mathbf{\tilde C_{\ell}^{th}}|} - \ln \left( \frac{|\mathbf{\tilde C_{\ell}^{th}}|}{|\mathbf{\tilde C_{\ell}^{fid}}|} \right)-n\right ]
\end{equation}
where $\mathbf {\tilde C_{\ell}^{th}}$ and $\mathbf {\tilde C_{\ell}^{fid}}$ are defined in equation~\eqref{clobs}, and $n$ is the number of observables, i.e.\ $T$, $E$ and $\Phi$ for Planck. The quantity $D_\ell$ is defined as follows:
\begin{equation} \label{D}
\begin{split}
D_{\ell} = \tilde C_{\ell}^{\mathrm{th},TT} \tilde C_{\ell}^{\mathrm{th},EE} \tilde C_{\ell}^{\mathrm{fid},\Phi \Phi} + \tilde C_{\ell}^{\mathrm{th},TT} \tilde C_{\ell}^{\mathrm{fid},EE} \tilde C_{\ell}^{\mathrm{th},\Phi \Phi}+ \tilde C_{\ell}^{\mathrm{fid},TT} \tilde C_{\ell}^{\mathrm{th},EE} \tilde C_{\ell}^{\mathrm{th},\Phi \Phi} \\ - \tilde C_{\ell}^{\mathrm{th},TE} \left( 2\tilde C_{\ell}^{\mathrm{fid},TE} \tilde C_{\ell}^{\mathrm{th},\Phi \Phi}+\tilde C_{\ell}^{\mathrm{th},TE} \tilde C_{\ell}^{\mathrm{fid},\Phi \Phi} \right)
\end{split}
\end{equation}
To obtain this expression we assume that each $a_{\ell m}$ of the maps follows Gaussian statistics and that there is no coupling between modes. Complications such as masks and anisotropic noise are neglected.

\subsubsection{CMB-S4}

\begin{table}[t]
\centering
\begin{tabular}{|c |c c| c| }
 \hline
 \multicolumn{4}{ |c| }{CMB-S4} \\
 \hline
 Channel [GHz] & \multicolumn{2}{ |c| }{ Noise [$\mu$K$^2$]} & Beam FWHM [arcmin] \\
 &Temperature  & Polarization&  \\
  \hline
\multicolumn{4}{ |c| }{SAT} \\
  \hline
145&2.13e-06&8.53e-06&25.5\\
155&4.13e-06&1.65e-05&22.7\\
220&1.32e-05&5.27e-05&13.0\\
270&3.87e-05&1.55e-04&13.0\\
\hline
\multicolumn{4}{ |c| }{LAT  (Chile)} \\
  \hline
145&1.22e-06&4.86e-06&1.4\\
155&1.16e-05&4.62e-05&1.0\\
220&7.20e-05&2.88e-04&0.9\\
\hline
\multicolumn{4}{ |c| }{LAT (South Pole)} \\
  \hline
145&2.43e-06&4.86e-06&1.4\\
155&2.89e-05&4.62e-05&1.0\\
220&1.80e-04&2.88e-04&0.9\\
 \hline
\end{tabular}
\caption{\label{CMB-S4} We summarize in this table the characteristics of the CMB-S4 configuration with 4 instruments, where the 2 LAT in Chile are considered as one instrument with twice the number of detectors. The effective survey time is $10\%$ of 5 years. These characteristics are taken from \cite{Abazajian:2019eic}.}
\end{table}
For completeness' sake, we extend our analysis to the future ground-based experiment CMB-S4 described in \cite{Abazajian:2016yjj,Abazajian:2019eic}. This survey will in particular improve the observations at high-$\ell$.
The current CMB-S4 proposal consists of 4 instruments: 
\begin{itemize}
    \item 3 Large-Aperture Telescopes (LAT) which are able to access high multipoles thanks to a very small beam, but are limited by atmospheric noise at low-$\ell$. The range of multipoles is assumed to be $[1000, 5000]$. 
    \item 1 Small-Aperture Telescope (SAT) which has low noise at low-$\ell$ but a large beam. The range of multipoles is assumed to be $[30, 1000]$. 
\end{itemize}
These telescopes will be shared between the South Pole and Chile. At the South Pole, one SAT and one LAT will be installed to observe one single patch of $3\%$ of the sky, since a small and deep patch is needed to detect a small value of $r$. The LAT is useful to have access to high multipoles for de-lensing. In Chile, 2 LAT will be installed to have access to almost $70\%$ of the sky ($60 \%$ after the galactic cut) and very high multipoles in order to achieve a high accuracy on the effective number of neutrino species $N_\mathrm{eff}$. The high-multipole measurement will allow us to reconstruct the lensing up to at least $\ell\sim1000$. Therefore, as for LiteBIRD, we will suppose a flat lensing noise \eqref{NP} on large scales up to $\ell=1000$ and neglect the information at larger multipoles. We use the configuration given in table~\ref{CMB-S4} for our analysis.\par

In addition to the usual white noise, we have to consider the atmospheric noise, which limits the measurements at low-$\ell$. Following \cite{Ade:2018sbj}, we model the noise as the usual white noise plus a contribution coming from the atmosphere: 
\begin{equation}
\label{atm}
n_{\ell}^{\lambda_1\lambda_2} = n_{\rm white}^{\lambda_1\lambda_2}+N_{\rm red}^{\lambda_1\lambda_2} \left(\frac{\ell}{\ell_{\rm knee}^{\lambda_1\lambda_2}} \right)^{\alpha_{\rm knee}^{\lambda_1\lambda_2}}
\end{equation}
where the subscript $\rm red$ means that we expect red noise from the atmosphere, i.e. with $\alpha_{\rm knee}$ negative. Recall that $n_{\ell}^{\lambda_1\lambda_2}$ is diagonal. As in \cite{Abazajian:2019eic} and given \cite{Ade:2018sbj}, we take $\ell_{\rm knee}=55$ and $\alpha_{\rm knee}=-2.5$ for both polarization and temperature in the case of the SAT. Actually, the temperature measurements of the SAT do not bring additional constraints for our purposes, since Planck temperature measurements in the SAT $\ell$-range are already almost cosmic variance limited. Regarding the LAT, for temperature we take $\ell_{\rm knee}=1000$ and $\alpha_{\rm knee}=-3.5$ and for polarization we take $\ell_{\rm knee}=700$ and $\alpha_{\rm knee}=-1.4$. In general we assume $N_{\rm red}^{\lambda_1\lambda_2}=n_{\rm white}^{\lambda_1\lambda_2}$, where $n_{\rm white}^{\lambda_1\lambda_2}$ is the amplitude of the white noise given in table~\ref{CMB-S4}, except for LAT temperature where we take $[9.51,108,196]\times 10^{-5}\mu$K$^2$, respectively, for the three LAT channels.
\par
Since we do not have access to the full sky from the ground, the power spectrum measurements are correlated between different $\ell$. We then bin the power spectra, since the typical correlation length is $\Delta \ell \sim 1/f_{\rm sky}$. For each bin, we assume the values of the power spectra to be the mean values inside the bin. Thus, the likelihood given in \eqref{corelik} becomes:
\begin{equation} \label{S4}
\mathcal{L}(\boldsymbol{\theta},\alpha,\cos{\Delta}) = - \frac{1}{2} \sum_{i} \sum_{\ell \in i} f_{\rm sky} (2\ell+1) \left[ \frac{D_{i}}{|\mathbf{\tilde C^{th}}_i|} - \ln \left( \frac{|\mathbf{\tilde C^{th}}_i|}{|\mathbf{\tilde C^{fid}}_i|} \right)-n\right ]  
\end{equation}
where $i$ stands for the bin number and the quantities with subscript $i$ have been averaged over the bin $i$.

\subsection{Bispectrum likelihood}
It is impossible to calculate the full bispectrum for each multipole combination $\ell_1$, $\ell_2$ and $\ell_3$ because of the high cost of the operation. Thus we use the binned bispectrum estimator \cite{Bucher:2015ura, Bucher:2009nm} in which we average the bispectrum over ranges of $\ell$. This operation is feasible because the theoretical bispectra we are looking for have features typically on the scale of the acoustic peaks. The information we loose, $\sim 1\%$ with a very limited number of bins $\sim 50$, is very small and provides a huge gain in calculation time and memory. We change the indices $\ell$ to $i$ to express this binning, i.e.\  $b_{i_1 i_2 i_3}$ is the averaged value of the bispectrum over the intervals
labeled $i_1$, $i_2$, $i_3$ of $\ell_1$, $\ell_2$, $\ell_3$ values. We use a matched filter to estimate the amplitude of specific theoretical shapes in the observed bispectrum, see \eqref{bisp}. \par
There exists no exact likelihood for the bispectrum. However, we can construct the following estimator for the $\tilde{f}_{\rm NL}^{I,JK}$ bispectrum amplitude parameters defined in \eqref{bisp}:
\begin{equation}
\label{ideal}
    \hat {\tilde f}_{\rm NL}^{I,JK} = \frac{ \left< \tilde b^{I,JK}, \tilde b^\mathrm{obs} \right>}{\left< \tilde b^{I,JK}, \tilde b^{I,JK} \right>}
\end{equation}
where $\tilde b^\mathrm{obs}$ is the observed reduced bispectrum. Furthermore, we have defined the inner product:
\begin{equation}
\label{scalarproduct}
\left< b, b^{\prime} \right> =
\sum_{i_1 \leq i_2 \leq i_3}
\sum_{ \lambda_{\substack{1,2,3, \\ 4,5,6}} }
b_{i_{1,2,3}}^{\lambda_{1,2,3}} \left( V^{-1} \right) _{i_{1,2,3}}^{\lambda_{1,2,3,4,5,6}}
b^{\prime \lambda_{4,5,6}}_{i_{1,2,3}}
\end{equation}
where we list different indices just by enumerating the numbers. For example $\lambda_{1,2,3}$ means $\lambda_1,\lambda_2,\lambda_3$ (remember that the $\lambda$ are polarization indices). Exactly like in the power spectrum case, we use a tilde to indicate that the observed bispectrum is the true bispectrum times the beam transfer functions plus noise: $\tilde{b}^\mathrm{obs}  = h_{\ell_a}h_{\ell_b}h_{\ell_c}  b^\mathrm{obs} + n_{\ell_a \ell_b \ell_c}$ (indices of the bispectrum are implicit). For the theoretical bispectrum the tilde means that we have multiplied it by the beam transfer functions. The matrix $V$ is the variance of the observed bispectrum determined in the weak non-Gaussianity approximation, which depends on the power spectra, the beam transfer functions and the noise. The estimator \eqref{ideal} is nearly optimal only for rotationally invariant maps. However, for real observations rotational invariance is broken because of the mask and the non-uniform noise. We can restore the optimality of the estimator by subtracting a ``linear term'' (linear in $a_{\ell m}$) from the observed bispectrum:
\begin{equation}\label{lin}
    b^{\lambda_{1,2,3},\mathrm{obs}}_{i_{1,2,3}} \rightarrow b^{\lambda_{1,2,3},\mathrm{obs}}_{i_{1,2,3}} -b^{\lambda_{1,2,3},\mathrm{lin}}_{i_{1,2,3}}
\end{equation}
For more details about the estimator, see \cite{Ade:2013ydc,Bucher:2015ura, Bucher:2009nm}.\par
In principle, the theoretical bispectrum is a function of all the cosmological parameters and of the six different isocurvature $\mathbf{\tilde f_{\rm NL}}$. However, as shown in \cite{Liguori:2008vf}, the statistical estimation of the cosmological parameters $\boldsymbol{\theta}$ would have a significant impact on the $\mathbf{\tilde f_{\rm NL}}$ error bars only if the detected $\mathbf{\tilde f_{\rm NL}}$ would have large signal-to-noise, equivalent to the signal-to-noise of the cosmological parameters. This is why we fix  $\boldsymbol{\theta}$ to the best estimated values determined from the power spectra alone in the Planck 2018 analysis \cite{Aghanim:2018eyx}, $\boldsymbol{\theta}^0$, so that the theoretical bispectrum is now only a function of the  $\mathbf{\tilde f_{\rm NL}}$, even if we allow $\boldsymbol{\theta}$ to vary for the power spectrum in our joint analysis. Using the theoretical bispectra, we can also estimate the Fisher matrix $ \mathbf F$ given in \cite{Bucher:2015ura, Langlois:2012tm}, with components:
\begin{equation}
\label{fisher}
 F_{ij} = \left< \tilde{b}^{(i)} , \tilde{b}^{(j)} \right>
\end{equation}
where $(i)$ and $(j)$ are any of the six combinations $(\zeta, \zeta\zeta)$, $(\zeta, \zeta S)$, $(\zeta, SS)$, $(S, \zeta\zeta)$, $(S, \zeta S)$ or $(S, SS)$.\par
The PDF of the $\tilde f_{\rm NL}$ is estimated as being Gaussian. We can then reduce the bispectrum data to only 6 observables, the $\tilde{f}_\mathrm{NL}^{I,JK}$, by constructing an effective likelihood directly of the $\tilde f_\mathrm{NL}$, instead of using the bispectrum distribution. This has the huge advantage of saving a lot of computation time with negligible impact on the performance. We use \eqref{ideal} with the transformation \eqref{lin} to obtain the best estimated values $\mathbf{\tilde f_{\rm NL}^0}$ and use \eqref{fisher} to estimate the Fisher matrix. Estimations of $\mathbf{\tilde f_{\rm NL}^0}$ and $F_{ij}$ are model independent, in particular they do not use the relations \eqref{fnl}. We express the bispectrum likelihood as an effective six-dimensional Gaussian function of the $ \tilde f_{\rm NL}$:
\begin{equation} \label{probafnl}
-2 \ln P \left(\boldsymbol{\xi}\right) = \left( \mathbf {\tilde f_{\rm NL}\left(\boldsymbol{\xi}\right)} -\mathbf { \tilde f_{\rm NL}^0} \right)^T  \mathbf{F}  \left( \mathbf{\tilde f_{\rm NL}\left(\boldsymbol{\xi}\right)} - \mathbf {\tilde f_{\rm NL}^0} \right)
\end{equation}
where $\mathbf{\tilde f_{\rm NL}\left(\boldsymbol{\xi}\right)}$ is defined in equation \eqref{fnl}. This is a good approximation since the $\tilde f_\mathrm{NL}$ estimator \eqref{ideal} depends linearly on the observed bispectrum. The observed bispectrum is obtained from the product of three $a_{\ell m}$ and is not Gaussian. However, the bispectrum value in each bin is the result of the average over many multipoles $\ell$ and $m$, such that the Gaussianity can be ensured by the central limit theorem. The values $\mathbf {f_{\rm NL}^0}$ without tilde can be found in \cite{Akrami:2019izv}. We recall that the $\mathbf {f_{\rm NL}^0}$ are defined with respect to the gravitational potential $\psi$, while the $\mathbf {\tilde f_{\rm NL}^0}$ are defined with respect to the curvature perturbation $\zeta$.\footnote{The exact conversion factors are for $(\zeta, \zeta\zeta)$, $(\zeta, \zeta S)$, $(\zeta, SS)$, $(S, \zeta\zeta)$, $(S, \zeta S)$, $(S, SS)$: -6/5, -2/5, -2/15, -18/5, -6/5, -2/5, respectively.} The resulting values for 
$\mathbf {\tilde f_{\rm NL}^0}$ after conversion, as well as its error bars for different experiments, are given in table~\ref{errortable}.
 \begin{table}
 \centering
 \begin{tabular}{ | l l l |l l l l l l|}
  \hline
 &&&$\zeta,\zeta\zeta$&$\zeta,\zeta S$&$\zeta,SS$&$S,\zeta\zeta$&$S,\zeta S$&$S,SS$ \\
 \hline
 \hline
 CDM & $\mathbf {\tilde f_{\rm NL}^0}$& Planck  & $-4.8$& $5.6$& $413$& $-345$&$-228$&$256$\\
 \hline
 & $\Delta \mathbf {\tilde f_{\rm NL}}$ &Planck &12&8&207&181&221&162 \\
 \hline
 & $\Delta \mathbf {\tilde f_{\rm NL}}$& LiteBIRD &9&8&72&75&70&57 \\
 \hline 
 & $\Delta \mathbf {\tilde f_{\rm NL}}$& CMB-S4 &12&20&211&173&201&283 \\
 \hline
 \hline
 $\nu$ density&$\mathbf{\tilde f_{\rm NL}^0}$&Planck& $64$ & $-64$& $547$& $-158$&$-420$&$800$  \\
 \hline
 & $\Delta \mathbf {\tilde f_{\rm NL}}$& Planck &34&43&224&176&289&389  \\
 \hline
  & $\Delta \mathbf {\tilde f_{\rm NL}}$& LiteBIRD &20&25&107&87&113&155 \\
  \hline
  & $\Delta \mathbf {\tilde f_{\rm NL}}$& CMB-S4 &29&51&214&141&245&473 \\
 \hline 
 \hline
 $\nu$ velocity & $\mathbf {\tilde f_{\rm NL}^0}$& Planck  & $-2.4$&$-100$&$280$&$133$&$-28$&$-15$ \\
 \hline
 & $\Delta \mathbf {\tilde f_{\rm NL}}$& Planck &29&34&118&93&91&113\\
 \hline
  & $\Delta \mathbf {\tilde f_{\rm NL}}$& LiteBIRD &19&10&36&50&33&18\\
  \hline
  & $\Delta \mathbf {\tilde f_{\rm NL}}$& CMB-S4 &42&38&108&118&113&100\\
  \hline
 \end{tabular}
 \caption{We give in this table for each isocurvature mode the best estimated value of $\tilde f_{\rm NL}$ given the Planck results \cite{Akrami:2019izv} as well as the associated standard deviation. We also give error forecasts for LiteBIRD and CMB-S4. \label{errortable}}
 \end{table}

\subsection{Implementation}
To perform the analyses, we use the MCMC statistical method \cite{Lewis:2002ah,Lewis:2013hha} with different power spectrum likelihoods for different experiments: the Planck likelihood, the likelihood \eqref{corelik} for LiteBIRD, and \eqref{S4} for CMB-S4. We modified the code \textit{cobaya}\footnote{\url{https://ascl.net/1910.019}} \cite{Torrado:2020dgo}, which includes the advanced MCMC sampler $CosmoMC$ and allows to sample arbitrary priors and posteriors. Results are analyzed using GetDist \cite{Lewis:2019xzd}. We generate all the power spectra by calling CAMB\footnote{\url{https://camb.info/}} \cite{Lewis:1999bs,Howlett:2012mh} twice in order to make linear combinations, because CAMB can only calculate the power spectra for total positive or negative correlation, i.e.\ $\cos \Delta = \pm1$. If we call the totally correlated power spectrum $C_{\ell}^{\lambda_1\lambda_2,+}$ and the totally anti-correlated power spectrum $C_{\ell}^{\lambda_1\lambda_2,-}$, we can compute the power spectrum for the case of an arbitrary correlation as follows:
\begin{equation}
\label{Clcamb}
C_{\ell}^{\lambda_1\lambda_2}= \frac{1}{2}\left( C_{\ell}^{\lambda_1\lambda_2,+}+C_{\ell}^{\lambda_1\lambda_2,-} + \cos \Delta \left(C_{\ell}^{\lambda_1\lambda_2,+}-C_{\ell}^{\lambda_1\lambda_2,-} \right)\right)
\end{equation}
Then \textit{cobaya} calls the relevant power spectrum likelihood.\par 

To perform a joint analysis that includes the information from the bispectrum, we first estimate $\tilde f_{\rm NL}^0$ and the associated Fisher matrix as described before to be able to compute \eqref{probafnl}. The joint analysis is performed by multiplying the power spectrum and bispectrum likelihoods, see equation \eqref{joint}. 

\section{Results}\label{result}
This section contains the results of our analyses.
In section \ref{Planck joint analysis}, we perform a joint analysis of the Planck power spectrum and bispectrum assuming two different cases: fixing $\cos \Delta$ or fixing $\kappa$, as we will see that when all parameters are left free, the joint analysis does not improve constraints. In section \ref{Theoretical results}, we discuss and summarize the usefulness of the joint analysis for many possible configurations using theoretical arguments. Finally, in section \ref{future} we compute forecasts for future experiments. We first investigate the possibility to detect isocurvature modes and their non-Gaussian features in these experiments. We then show the result of the joint analysis in the favourable cases.\par
In this section, we will always show results for $\beta_{\rm iso}$ instead of $\alpha$. Recall that $\beta_{\rm iso}=\alpha/(1+\alpha)$. It is convenient for the analysis to use $\beta_{\rm iso}$ because it is bounded between 0 and 1. Moreover, all the results of Planck are given in terms of $\beta_{\rm iso}$. However, for the small values of $\beta_{\rm iso}$ allowed by the power spectrum, we can say that $\alpha \approx \beta_{\rm iso}$. Recall that we apply a flat prior on $\mathcal P_{IJ}^{(1)}$ and $\mathcal P_{\zeta \zeta}^{(2)}$ defined in \eqref{primordial}.  \par

\subsection{Planck joint analysis} \label{Planck joint analysis}
\begin{figure}[t]
\centering
\includegraphics[scale=0.19]{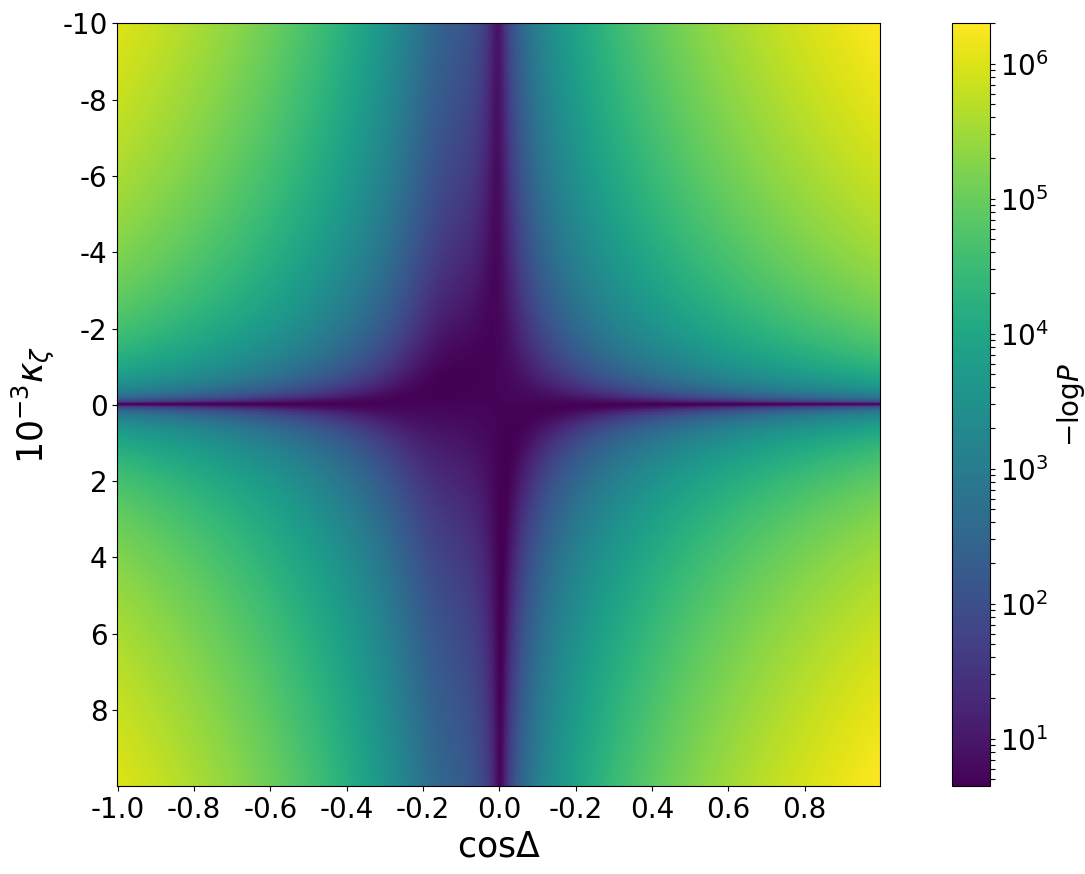}
\includegraphics[scale=0.19]{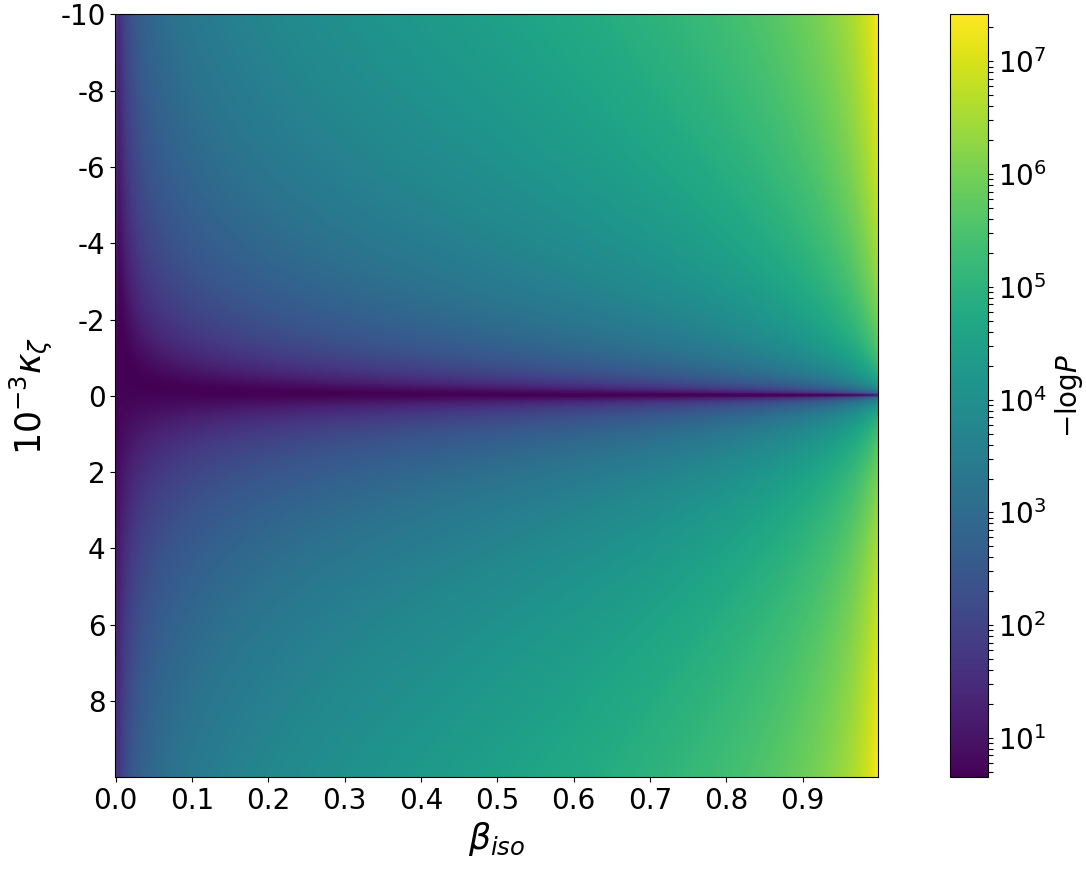}
\caption{\label{bispectre} CDM isocurvature bispectrum PDF, $-\ln P$ defined in (\ref{probafnl}), as a function of $\cos\Delta$ and $\kappa_{\zeta}$ with $\beta_{\rm iso}=0.016$ and $\kappa_{S}=0$ in the panel on the left, and of $\beta_{\rm iso}$ and $\kappa_{\zeta}$  with $\cos \Delta=-0.1$ and $\kappa_{S}=0$ in the panel on the right. }
\end{figure}
The joint analysis, given our model, does not improve constraints in general in the case of Planck, i.e.\ without detection of isocurvature modes in the power spectrum and without detection of primordial non-Gaussianity in the bispectrum. We can directly see this from figure~\ref{bispectre}. There is a strong degeneracy between the parameters $\kappa_I$ in the bispectrum and the power spectrum parameters $\beta_{\rm iso}$ and $\cos\Delta$. For total (anti-)correlation, i.e.\ $\cos \Delta=\pm1$, we have well constrained $\kappa_I$ which in that case are directly linked to $\tilde f_{\rm NL}^{I,\zeta \zeta}$. However, for $\cos \Delta$ close to and compatible with 0, as the power spectrum constraints that we gave in section~\ref{Planck likelihood} tell us, the parameters $\kappa_I$ can take arbitrarily large values as we see in the left panel of figure~\ref{bispectre} for $\kappa_\zeta$. In principle the 1$\sigma$ and 2$\sigma$ contours should go to infinity, but for very large $\kappa_I$, the width in the $\cos \Delta$ dimension becomes very small so that it becomes difficult to sample.
The right panel of figure~\ref{bispectre} is similar, but this time as a function of $\beta_{\rm iso}$ instead of $\cos\Delta$. Again, for $\beta_{\rm iso}$ close to and compatible with 0, as given by the power spectrum constraints, $\kappa_{\zeta}$ can take arbitrarily large values, so the space to sample in this direction is infinite.  \par
The previous paragraph concerned the bispectrum analysis alone. If we add the power spectrum constraints, only the measurements of $\cos \Delta$ and $\beta_{\rm iso}$ will be improved since the power spectrum does not depend on the $\kappa_I$. The constraints are compatible with 0 for both the isocurvature amplitude and the correlation to a high probability, thus the remaining space to sample is again infinite. One could integrate numerically over the $\kappa_I$ and obtain constraints on $\beta_{\rm iso}$ and $\cos \Delta$. However, these constraints would basically be meaningless because they depend completely on the chosen parametrisation and are independent of the Fisher matrix.
In other words, the $\kappa_I$ absorb all the constraints from the bispectrum and since the power spectrum does not depend on $\kappa_I$, the constraints on the other parameters are not improved.\par
For the joint analysis with the bispectrum to have any effect for Planck, we have to fix some of the parameters. Some models are able to predict $\cos \Delta=\pm1$, and we assume it might be possible to have models that predict other non-zero values as well. Fixing the correlation is equivalent to choosing as bispectrum likelihood a slice of constant $\cos \Delta$ in the left panel of figure~\ref{bispectre}.  Other theoretical models might have specific predictions for the $\kappa_I$ parameters.
\begin{figure}[t]
\centering
\includegraphics[scale=0.22]{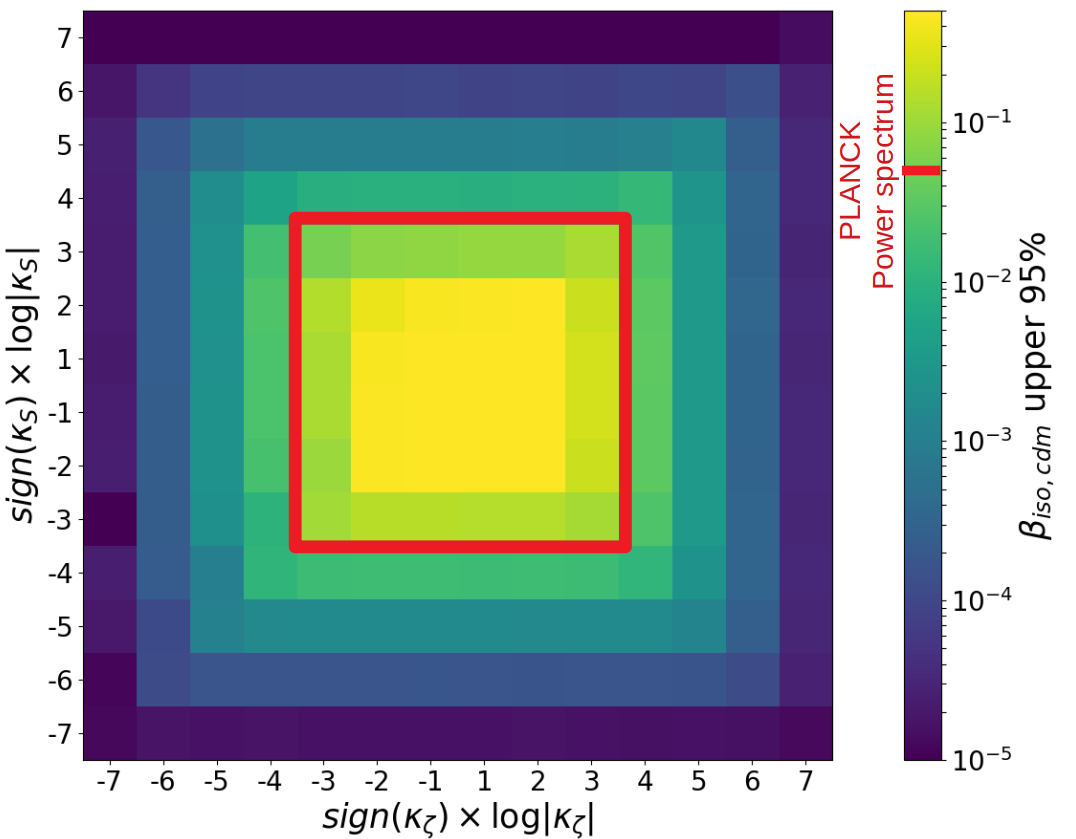}
\includegraphics[scale=0.45]{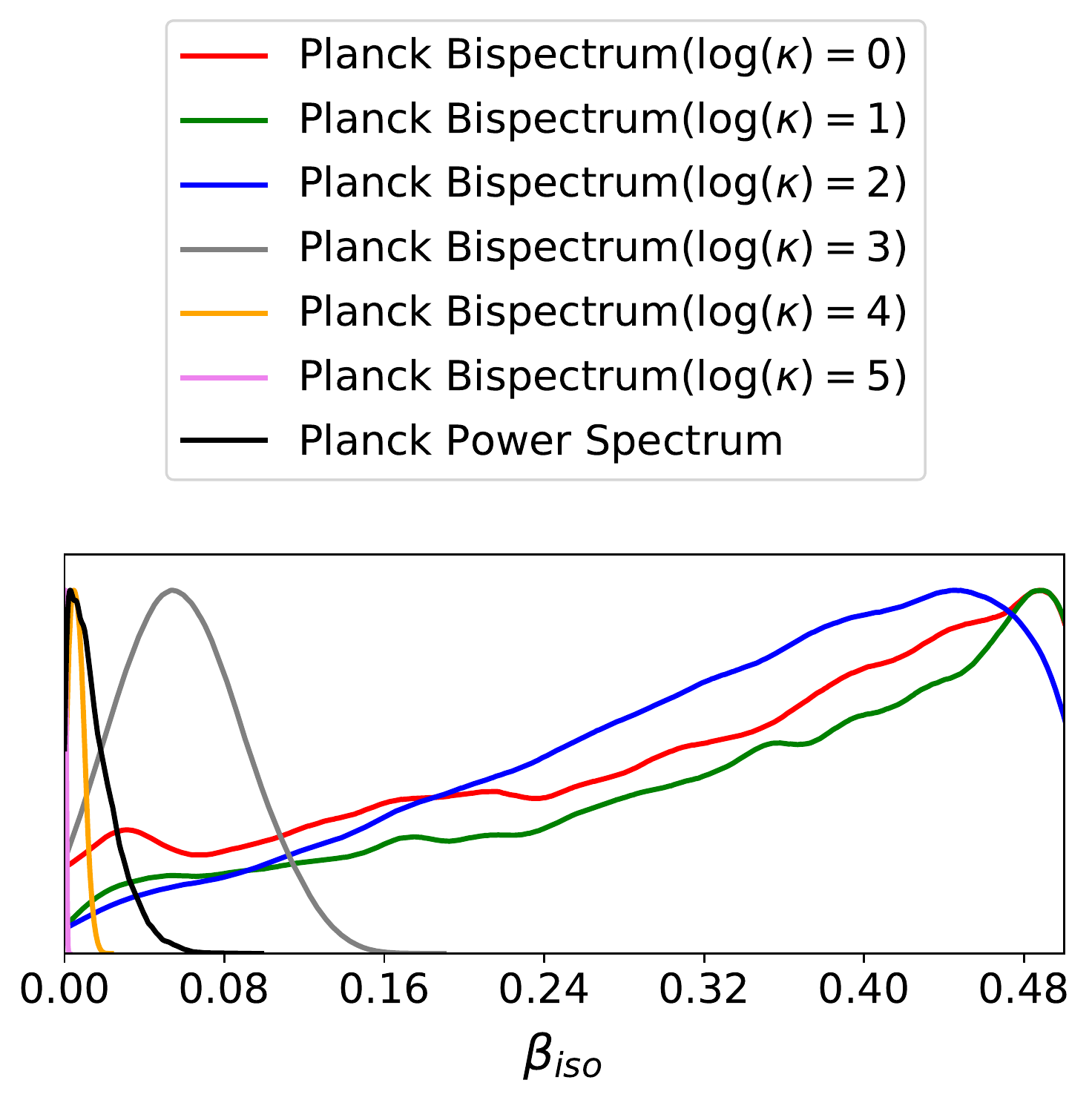}
\caption{\label{kappakappa} In the left panel, we show the 2$\sigma$ upper bound of the parameter $\beta_{iso,cdm}$, marginalizing over $\cos\Delta$, as a function of the two parameters $\kappa_{\zeta},\kappa_S$. The upper bound is calculated with a MCMC using only the bispectrum likelihood of equation \eqref{probafnl} and fixing the couple $\kappa_{\zeta},\kappa_S$. The red square is the 95\% upper bound given by the Planck power spectrum. In the right panel, except for the black curve, we show the probability distribution for the specific case where $\kappa_{\zeta} = \kappa_{S} $. So the curves of the right panel correspond to the positive diagonal in the left panel. We also plot the PDF obtained with the power spectrum alone (black curve).}
\end{figure}

\subsubsection{General correlation, fixed $\kappa_I$}\label{General correlation}
 
\paragraph{Constraints on $\beta_{\rm iso}$:}
We assume here a model where the parameters $\kappa_I$ are predicted by theory. In the model studied in \cite{Langlois:2012tm,Langlois:2011hn,Langlois:2011zz}, where the curvaton decays into CDM and radiation, the parameters $\kappa_I$ can be expressed as a function of two parameters $f_c$ and $r$. The former represents the fraction of CDM created by the decay and the latter quantifies the transfer between the pre-decay and post-decay perturbations. It is not hard to imagine that in some specific particle theory those quantities could be computed and hence the values of the $\kappa_I$ would be fixed by theory. In equation \eqref{fnl}, we see that $\cos^2\Delta$, $\beta_{\rm iso}$ and combinations of the two depend on $\tilde f_{\rm NL}/\kappa_I$ directly. Two parameters are fixed: for $I=\zeta$ and for $I=S$. In figure \ref{kappakappa}, we show the constraints from the bispectrum by representing the 2$\sigma$ upper value of $\beta_{iso,cdm}$ obtained with a MCMC chain marginalized over $\cos\Delta$ using only \eqref{probafnl} as a function of the chosen values of $\kappa$. The flat priors of $\mathcal P^{(1)}_{SS}$ and  $\mathcal P^{(1)}_{\zeta \zeta}$ have the same upper limit, so that we can have the same amplitude in the adiabatic and in the CDM isocurvature mode. It corresponds to $\alpha=1$ and $\beta_{\rm iso}=0.5$.
We should have no constraint on $\alpha$ coming from the bispectrum for small $\kappa$, so that we find the same posterior distribution as our prior which is flat for $\mathcal P^{(1)}_{SS}$ and hence $\alpha$. The change of variable $\alpha \rightarrow \beta_{\rm iso}$ contracts intervals of $\alpha$ and hence makes higher $\beta_{\rm iso}$ values more likely. This happens for $\log \kappa = 0,1,2$. When $\log \kappa \geq 3$, the bispectrum provides additional constraints on the isocurvature amplitude. The larger $\kappa$ is, the more the bispectrum constrains $\beta_{\rm iso}$. It provides better constraints than the power spectrum for $\log \kappa \geq 4$.
\par
Hence we predict that the joint analysis will be able to improve the constraints for the models that are outside the red square in figure~\ref{kappakappa}. To make the analysis simpler and because only the largest $\kappa$ is relevant, we can set $\kappa_{\zeta}=\kappa_{S}=\kappa$ and only study the joint analysis for the diagonal of that figure.
\begin{figure}[t]
\centering
\includegraphics[scale=0.4]{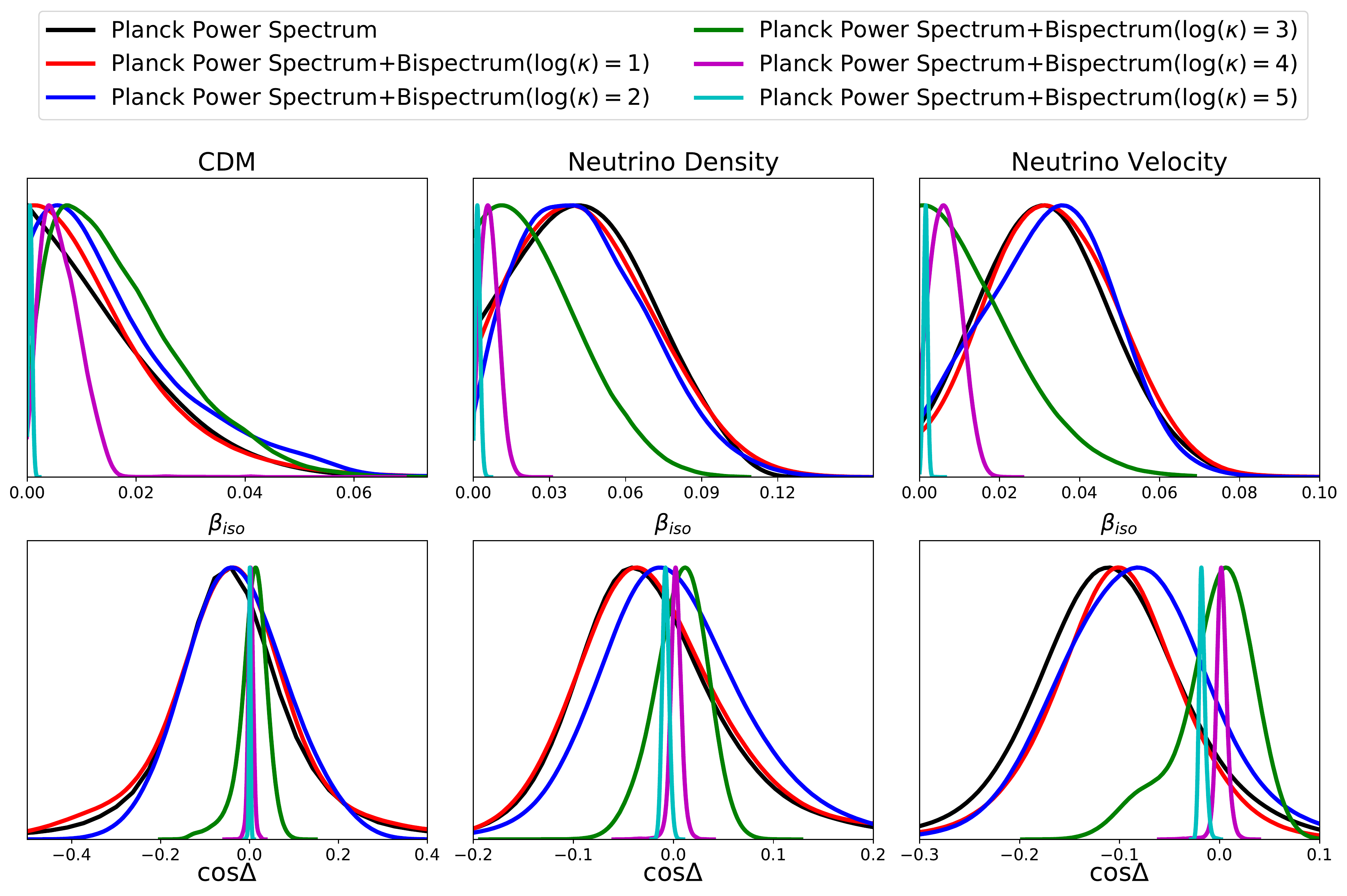}
\caption{Marginalized constraints of $\beta_{\rm iso}$ and $\cos \Delta$ for the joint analysis with fixed $\kappa$. Respectively, the first, second and third column correspond to CDM, neutrino density and neutrino velocity. The black curve shows the constraint obtained using the power spectrum alone. \label{planckcdm}}
\end{figure}
\par
\begin{figure}[t]
\centering
\includegraphics[scale=1.05]{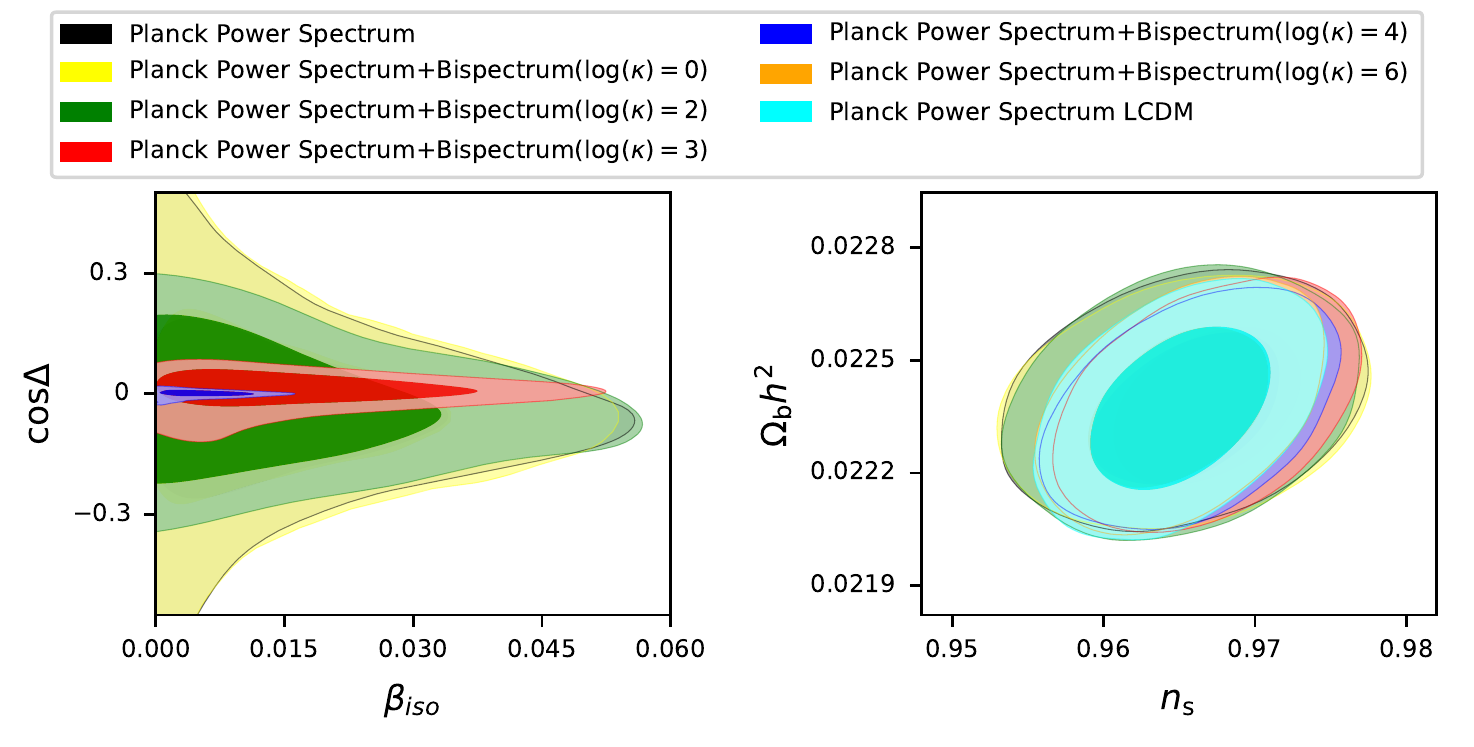}
\caption{\label{planckcdmother} Left: constraints of the joint analysis and of the power spectrum alone in the ($\cos \Delta $, $ \beta_{\rm iso}$) space for CDM isocurvature modes. Right: result of this analysis in ($\Omega_b$, $n_s$) space, i.e.\ constraints from the power spectrum alone compared with the joint analysis at different fixed values of $\kappa$. We also show the constraints given by the power spectrum in the case of the standard $\Lambda$CDM model, i.e.\ without isocurvature modes. For low values of $\kappa$, the joint analysis ($\Omega_b$, $n_s$) contour and the power spectrum contour assuming a model with one isocurvature mode are similar. For large $\kappa$, the joint analysis ($\Omega_b$, $n_s$) contour is similar with the contour given by $\Lambda$CDM without isocurvature mode. This behaviour of the constraints with respect to the joint analysis is similar for all other cosmological parameters. }
\end{figure}

 In figure \ref{planckcdm} we see that for large enough $\kappa$, values of $\cos\Delta$ and $\beta_{\rm iso}$ compatible with data are close to zero for all isocurvature modes. This result is consistent with what we observed in figure \ref{kappakappa} (right panel). The joint analysis starts to be efficient when the bispectrum constraints become comparable to the power spectrum constraints, represented by the black curve in figure \ref{kappakappa}.
In figure \ref{planckcdm}, for $\log \kappa = 1,2$ the constraints on the isocurvature parameters obtained with the joint analysis are comparable with the power spectrum constraints. We observe that the size of the $\beta_{\rm iso}$ contour increases  slowly to allow higher $\beta_{\rm iso}$ for the CDM isocurvature mode. Finally for $\log \kappa = 3,4,5$ contours are contracted near $0$. The intermediate behaviour where the contour increases slowly for $\log \kappa = 1,2$ can be understood by looking at the ($\beta_{\rm iso}$, $\cos \Delta$) space for the CDM isocurvature mode in the left panel of figure \ref{planckcdmother}. The power spectrum allows values of $\cos \Delta$ from $0.3$ to $-0.4$ at $2\sigma$ level. When $\kappa$ increases and the bispectrum constraints start to have an impact, regions where $|\cos \Delta|>0.3$ start to be excluded, see also the $\cos \Delta$ panel of figure~\ref{planckcdm}. The posterior distribution then can include higher values of $\beta_{\rm iso}$ due to the renormalization of the distribution. We have reproduced the same effect by imposing a prior with bounds $0.2$ and $-0.2$ for the correlation.
\par
Regarding the cosmological parameters, we can see in the right plot of figure \ref{planckcdmother} that for $\log\kappa = 0$, the contours of the joint analysis in the ($\Omega_b$, $n_s$) space are similar to contours obtained from the power spectrum alone including isocurvature modes. This means that the bispectrum does not constrain the isocurvature modes for $\kappa$ too small. However, when $\log \kappa \geq 5,6$, the contours of $\Omega_b$ and $n_s$ are equivalent to the contours given by the power spectrum alone assuming no isocurvature modes. In general, the contours of the cosmological parameters are not degraded by the estimation of the isocurvature mode parameters if $\kappa_I$ is large. The other parameters of $\Lambda$CDM have the same behaviour. We understand this result because when $\kappa$ tends to infinity, the constraints on the $\tilde f_{\rm NL}$ from the bispectrum provide very tight constraints on $\beta_{\rm iso}$ and $\cos \Delta$ given \eqref{fnl}. In other words, setting $\log \kappa \geq 6$ gives the same constraints on the parameters as a purely adiabatic model. While figure~\ref{planckcdmother} is for CDM isocurvature, results are similar for the other isocurvature modes.\par

\paragraph{Constraints on $\tilde f_{\rm NL}$:}
Fixing $\kappa$ can improve the power spectrum constraints on $\beta_{\rm iso}$ and $\cos\Delta$ if $\kappa > 10^3$. However, a fixed $\kappa$ in combination with the constraints on $\beta_{\rm iso}$ and $\cos\Delta$ also allows us to derive constraints on the $\tilde f_{\rm NL}$ parameters, thanks to \eqref{fnl}. In the range where the bispectrum does not improve $\beta_{\rm iso}$ and $\cos \Delta$, i.e.\ when $\kappa<10^3$, we might even say that the $\tilde f_{\rm NL}$ error bars are strongly improved since the $2\sigma$ ranges are $\tilde f_{\rm NL}^{I,\zeta \zeta}<0.18$, $-0.03<\tilde f_{\rm NL}^{I,\zeta S}<0.02$ and $\tilde f_{\rm NL}^{I,SS}<0.04$ for the case where $\kappa=1$. In the range where the bispectrum improves the constraints we obtain the following $2\sigma$ ranges: $\tilde f_{\rm NL}^{I,\zeta \zeta}<2\times10^{-4}$, $-6\times10^{-4}<\tilde f_{\rm NL}^{I,\zeta S}<9\times10^{-4}$ and $\tilde f_{\rm NL}^{I,SS}<0.01$ for the case where $\log(\kappa)=4$. 
Indeed, the equations  \eqref{fnl} give the $\tilde f_{\rm NL}$ as products of $\kappa$ with $\beta_{\rm iso}$ (or $\alpha$) and $\cos \Delta$, which are constrained to very small values by the power spectrum. However, we must be careful with this interpretation, since the $\tilde f_{\rm NL}$ error bars obtained in the usual Planck analysis are model independent, while we have assumed here a model which implies relations between the $\tilde f_{\rm NL}$. Furthermore, we have assumed a flat prior on $\boldsymbol{\xi}$ (defined above \eqref{joint}), while we would obtain the same results as the Planck analysis if we used a flat prior on the $\tilde f_{\rm NL}$ (and did not have those relations between the $\tilde f_{\rm NL}$). Hence our constraints are not directly comparable with the published Planck results regarding the $\tilde f_{\rm NL}$. 

\subsubsection{Fixed correlation, general $\kappa_I$}

\paragraph{Constraints on $\beta_{\rm iso}$:}
In this section, we assume a model where the correlation between the isocurvature mode and the adiabatic one is predicted. On the other hand, the $\kappa_I$ are now free parameters. Fixing $\cos \Delta$ can be seen as a re-scaling of $\tilde f_{\rm NL}$, but contrary to the previous case of fixed $\kappa$, each $\tilde f_{\rm NL}^{I,JK}$ is not re-scaled by the same factor, see \eqref{fnl}. We can also deduce from the formulas  \eqref{fnl} that $\kappa_I$ and $\cos \Delta$ will have opposite effects on the $\beta_{\rm iso}$ distribution: given the first equation involving $\tilde f_{\rm NL}^{I,\zeta \zeta}$, we see that fixing $\kappa$ to a high value is equivalent to fixing $\cos \Delta$ to a small value. We first show the result for $\cos \Delta=-1$, since this total anti-correlation can be theoretically motivated by a curvaton scenario as described for the first time in \cite{curvaton,curvaton1}. Then we take $\cos \Delta = -0.4$, which is the $2\sigma$ bound from the power spectrum for the CDM isocurvature mode \cite{Akrami:2018odb}, and finally two smaller values on a log-scale: $\cos \Delta = -0.1,-0.01$.  As in the case of fixed $\kappa$, negative and positive values give very similar results. We choose here a negative correlation since that is more likely for every isocurvature mode given the Planck constraints. \par
\begin{figure}[t]
\centering
\includegraphics[scale=0.37]{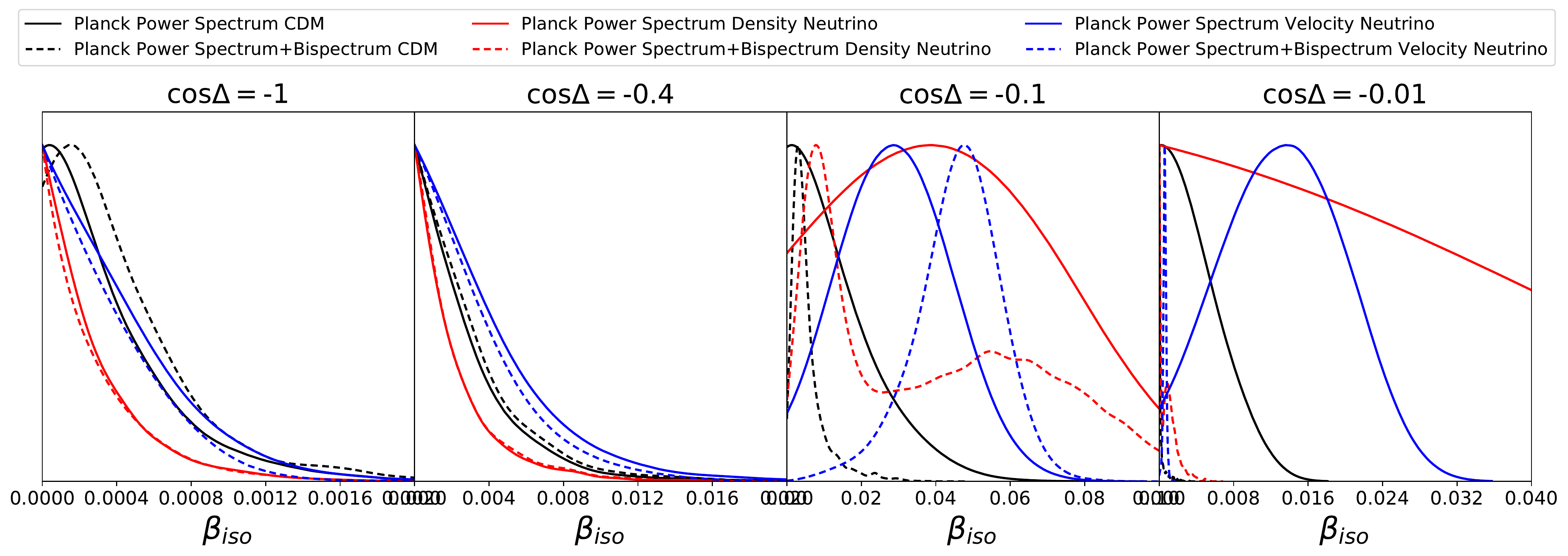}
\caption{Constraints obtained on $\beta_{\rm iso}$ when fixing $\cos\Delta$. We show for every isocurvature mode (CDM in black, neutrino density in red, and neutrino velocity in blue) the result of the power spectrum analysis alone (solid curve) and of the joint analysis with the bispectrum (dotted curve).  \label{cosR}}
\end{figure}

In figure \ref{cosR}, we see that for $\cos \Delta = -1,-0.4$ there is no improvement of the constraints compared to the power spectrum constraint alone, as expected since it is equivalent to small values of $\kappa_I$. In the range of correlations $\cos \Delta = -0.1,-0.01$, the bispectrum induces a contraction effect of more than $1\sigma$ for CDM and neutrino density. In the case of neutrino velocity, the bispectrum has pushed $\beta_{\rm iso}$ to larger values. This mode has the advantage to have two signal-to-noise values larger than $2$ in the model-independent bispectrum analysis: $2.9\sigma$ for $\tilde f_{\rm NL}^{\zeta,\zeta S}$  and $2.3\sigma$ for $\tilde f_{\rm NL}^{\zeta,SS}$. The ratio of the two can then directly constrain $\beta_{\rm iso}$ (through $\alpha$). For $\cos\Delta=-0.1$, we obtain from the bispectrum alone a central value of $0.07$ for $\beta_{\rm iso}$ for this mode, which is larger than the constraint of the power spectrum. The detection of this mode in this configuration is improved to $4.0\sigma$ thanks to the bispectrum. For $\cos\Delta=-0.01$, the ratio gives a central value for $\beta_{\rm iso}$ of about $8\times 10^{-4}$ (i.e.\ from the bispectrum alone), which leads to a central value of $6\times 10^{-4}$ for the joint analysis. Again, for the neutrino velocity mode, the ``detection'' of the two $\tilde f_{\rm NL}$ improves the detection of $\beta_{\rm iso}$ to $3.5\sigma$ for this configuration (which cannot be seen in figure~\ref{cosR} because of the scale). \par

At this point, we could claim that in the case of a model predicting a correlation of order $-0.1$ or $-0.01$, the joint analysis is able to detect $\beta_{\rm iso}$ for the neutrino velocity isocurvature mode in the Planck data. This result should be taken with care for the following reasons. First, our signal-to-noise values for the $\tilde f_{\rm NL}$ are slightly different from the ones given in \cite{Akrami:2019izv} because we use the Fisher error bars given in table~\ref{errortable} while \cite{Akrami:2019izv} computes the error bars from simulations. The Fisher error bars are smaller in the case of the neutrino velocity mode compared with the true error bars given in \cite{Akrami:2019izv} for $\tilde f_{\rm NL}^{\zeta,\zeta S}$  and $\tilde f_{\rm NL}^{\zeta,SS}$, which increases their signal-to-noise (recall also that we use in this paper $\tilde f_{\rm NL}$ defined in terms of $\zeta,S$ instead of $f_{\rm NL}$ defined in terms of the gravitational potential). Second, as discussed in \cite{Akrami:2019izv}, having one signal-to-noise larger than $2.5$ cannot be considered a detection given the large number of parameters measured and the lack of consistency between the temperature-only and the temperature+polarization results. Hence these might very well be simple statistical fluctuations, and basing any conclusions on them is risky. However, \cite{Akrami:2019izv} does not discuss the probability of having two signal-to-noise values larger than 2 in the same mode.

In figure~\ref{cosR}, we observe for neutrino density a second bump for a correlation of $-0.1$ and, although it cannot be seen because of the scale, there is actually also a similar second bump for $\cos\Delta=-0.01$. This is due to the strong correlation between $\kappa_I$ and $\beta_{\rm iso}$.
For all modes, when $\cos \Delta$ goes to 0, the isocurvature amplitude space also goes to 0. For a very small correlation, $\kappa_I$ could take any large value up to infinity and, as we saw in the case of fixed $\kappa$, the results in the $\Lambda$CDM parameter space then tend to what one would get with purely adiabatic initial conditions. This explains the first peak. When $\kappa_I$ becomes smaller, the power spectrum dominates the constraints, which gives us the second bump.

\paragraph{Constraints on $\tilde f_{\rm NL}$:}
As in section \ref{General correlation}, we can obtain constraints on $\tilde f_{\rm NL}$ as derived parameters. But exactly as before, we believe that those constraints are trivial given the relations \eqref{fnl} and they are prior dependent. Furthermore, in this particular case where we fix the correlation between the adiabatic and the isocurvature mode,  the infinite degeneracy between $\kappa$ and $\beta_{\rm iso}$ makes the marginalized constraints on $\tilde f_{\rm NL}^{I,\zeta\zeta}$ even weaker when we add the constraints of the power spectrum to the bispectrum. For all these reasons, we believe that these constraints are not meaningful.

\subsection{Theoretical assessment} \label{Theoretical results}
In this section, we give theoretical arguments to justify the choices we made in section \ref{Planck joint analysis} and to prepare the investigation of what will be possible with future experiments as described in the next section. More generally, we will study for all different possible cases of detection/non-detection and fixed/free parameters what we expect regarding the impact of a joint analysis of the power spectrum and the bispectrum.
\par
As we pointed out before, the six relations \eqref{fnl} for the $\tilde f_{\rm NL}$ parameters are not independent. Only 3 of them are independent and these are expressed in terms of 4 parameters. Unfortunately, the parameters $\kappa_I$ are degenerate with both $\cos \Delta$ and $\alpha$. This means in general that if we do not detect isocurvature modes, no constraints can be established on the $\kappa_I$. A detection, however, can break the degeneracy. Thus, our results depend on the detection of $\alpha$, $\cos \Delta$ and $\tilde f_{\rm NL}$.
Alternatively, we might have models where some of the parameters have a predicted value. For example, there are models that predict the adiabatic and isocurvature modes to be fully (anti-)correlated. One could also imagine models in which the $\kappa_I$ parameters are predicted. \par
In table \ref{thtable} we give the conclusions for different cases of detection of $\alpha$, $\cos\Delta$ and $\tilde f_{\rm NL}$ and of fixing the parameters $\cos \Delta$ and $\kappa_I$. What we mean by "detection" in this section is that the value 0 is excluded by at least $4\sigma$. The symbol "$\times$" means the condition ($\tilde f_{\rm NL}$ and $\alpha$ being detected, $\kappa$ being fixed to a specific value) is not satisfied, while "\checkmark" means it is. As for $\cos\Delta$ we have to consider both detection and fixing, we are more explicit in that column. In the table we have put only the most important conclusions for each case; for more information, see the corresponding discussion in the rest of this section. \par
We will now discuss table \ref{thtable} line by line. To refer to a specific equation of \eqref{fnl}, we will just give for example the combination $(I,\zeta\zeta)$ to refer to the two equations involving $\tilde f_{\rm NL}^{\zeta,\zeta \zeta}$ and $\tilde f_{\rm NL}^{S,\zeta \zeta}$. 
We start with the cases where $\cos \Delta$ is a free parameter:
\begin{table}
 \begin{tabular}{|c||c | c c c |c|}
 \hline
  & $\cos \Delta$ & $\tilde f_{\rm NL}$ & $\alpha$ or $\beta_{\rm iso}$ & $\kappa$ fixed & Joint analysis improves constraints \\
 \hline
  \hline
 1 & not detected & $\times$ & $\times$  & $\times$ &  $\times$  \\
 \hline
 2 &""  & $\times$ & \checkmark  & $\times$ &  $\times$\\
 \hline
 3 & ""  & $\times$ & $\times$  & \checkmark &  if $\kappa_I$ large enough \\
 \hline
 4 & ""  & \checkmark &  $\times$ & $\times$ &  $\times$  \\
  \hline
  5 & ""  & \checkmark &  \checkmark & $\times$ & constraints/detections of $\cos \Delta$ and $\kappa_I$ \\
  \hline
 6 & detected  & $\times$ & \checkmark  & $\times$ & if $\cos\Delta$ small enough  \\
 \hline
 7 & ""  & \checkmark & \checkmark  & $\times$ &  constraints on $\cos \Delta$ and $\kappa_I$ \\
 \hline
 \hline
 8 & fixed $\neq 0$ & $\times$ & $\times$  & $\times$ &   if $\cos \Delta$ small enough \\
 \hline
  9 & "" & \checkmark & $-$  & $\times$ &  constraint/detection of $\alpha$   \\
 \hline
 10 & fixed $= 0$  & $-$ & $-$  & $\times$ & $\times$  \\
 \hline
 11 & "" & $\times$ & $\times$  & \checkmark &   if $\kappa \uparrow$, $\alpha \downarrow$   \\
 \hline
 12 & "" & \checkmark & $\times$  & \checkmark &   detection of $\alpha$   \\
 \hline
\end{tabular}
\caption{Summary of the usefulness of performing a joint analysis of the power spectrum and the bispectrum (compared to an analysis of the power spectrum alone) for all possible configurations of detection/non-detection and fixed/free parameters as explained in the main text. The table is divided into two parts: from line 1 to 7, the correlation $\cos\Delta$ is free and from line 8 to 12, it is fixed. In the right column we give a brief conclusion for each case; for more details, see the corresponding description in the main text. (A $-$ symbol means that the conclusion is independent of that choice.)\label{thtable}}
\end{table}

\begin{enumerate}
\item \label{point1} Here, $\cos \Delta$ is not detected in the power spectrum and there is neither detection of $\tilde f_{\rm NL}$ in the bispectrum nor detection of isocurvature modes in the power spectrum. We have $\cos \Delta$ and $\beta_{\rm iso}$ compatible with zero and then $\kappa_I$ can take any arbitrarily large value. Since $\kappa_I$ is strongly degenerate with $\cos \Delta$ and $\beta_{\rm iso}$, a joint analysis would give artificial constraints that are only due to parametrization effects as we can see in figure~\ref{bispectre}. Furthermore, the $\kappa_I$ are also compatible with zero with high probability. This means that if we only consider the bispectrum, $\alpha$ can also take any arbitrarily large value, so the constraints only come from the power spectrum.
\item \label{point2} If we have a detection of $\alpha$ from the power spectrum with a future experiment, the PDF of $\kappa$ is now constrained, since from the equation $(I,SS)$ in \eqref{fnl} we have $\kappa_I = \tilde f_{\rm NL}^{I,SS}/\alpha$. But $\kappa$ remains compatible with zero because we have no detection of $\tilde f_{\rm NL}$, so we see from the equations $(I,\zeta S)$ and $(I,\zeta \zeta)$ that no additional constraint can be imposed on $\alpha$ or even on $\cos \Delta$.

\item \label{point3} Next we study the effect of fixing the $\kappa_I$ without any detection. Just by looking at equation $(I,S S)$ we see immediately that the more $\kappa$ is fixed to a large value, the more the PDF of $\alpha$ will be contracted to 0 such that the product with $\kappa$ fits with the PDF of $\tilde f_{\rm NL}^{I,S S}$. Actually, the same conclusion can be drawn from a combination of equations $(I,\zeta S)$ and $(I,\zeta \zeta)$. The product $\kappa_I \cos^2 \Delta$ in equation $(I,\zeta \zeta)$ is constrained by the PDF of $\tilde f_{\rm NL}^{I,\zeta\zeta}$, so if we fix $\kappa_I$ to a large enough value, the square of $\cos \Delta$ can be very small while the product $\kappa_I \cos \Delta$ can still be very big in equation $(I,\zeta S)$, pushing $\alpha$ towards zero. 

\item  \label{point4} Let us now study the consequences of a detection of $\tilde f_{\rm NL}$. Of course it is possible to have all intermediate cases where just one or some $\tilde f_{\rm NL}$ are detected, but let us assume the ideal case where all the $\tilde f_{\rm NL}$ are detected. The system \eqref{fnl} has more parameters $(\alpha,\cos\Delta,\kappa_{\zeta},\kappa_{S})$ than independent $\tilde f_\mathrm{NL}$. Therefore, we cannot break the degeneracy between the parameters. Moreover, if $\alpha$ and $\cos \Delta$ are still compatible with 0, the $\kappa_I$ PDF is not bounded which would give results that are difficult to interpret even if ratios of $\tilde f_{\rm NL}$ with the same first index $I$ are defined and do not depend on $\kappa_I$. For example we have that $\tilde f_{\rm NL}^{I,\zeta\zeta}/\tilde f_{\rm NL}^{I,SS} = \cos ^2 \Delta / \alpha$.

\item \label{point5} Let us assume here that we detect $\alpha$ in the power spectrum. This avoids the problem of unbounded $\kappa_I$. We then just need to detect one $\tilde f_{\rm NL}^{I,SS}$ to determine the corresponding $\kappa_I$, which will then also be detected. Thanks to the relations $(I,\zeta \zeta)$ and $(I,\zeta S)$, we could in that case improve the constraints on $\cos\Delta$ and possibly even improve the detection of $\alpha$, depending on the accuracy of the $\tilde f_{\rm NL}$ measurements. If we also detect $\tilde f_{\rm NL}^{I,\zeta S}$ or $\tilde f_{\rm NL}^{I,\zeta \zeta}$, the detection of the correlation $\cos\Delta$ could also be performed with the joint analysis. If $\tilde f_{\rm NL}^{I,SS}$ is not detected, we can still have constraints from the bispectrum by detecting either the couple $\tilde f_{\rm NL}^{\zeta,\zeta \zeta},\tilde f_{\rm NL}^{\zeta,\zeta S}$ or $\tilde f_{\rm NL}^{S,\zeta \zeta},\tilde f_{\rm NL}^{S,\zeta S}$. This way we can determine and detect both $\kappa_I$ and $\cos\Delta$. The other relation imposed by our model might then allow improvements of constraints on $\alpha$ and $\cos\Delta$. As we can determine the $\kappa_I$ from the data in this case, there is no need to study the case where they are fixed as well. 

\item \label{point6} We can now study the case where the isocurvature modes are detected in the power spectrum as well as their correlation with the adiabatic mode ($\cos \Delta$=0 excluded). Here again we will have a bounded PDF for $\kappa_I$. In general, the first equations $(I,\zeta \zeta)$ lead to the smallest error bars. If the detected value of $\cos \Delta$ is small enough, the same effect that contracts the PDF of $\alpha$ to 0 in point \ref{point3} will here contract the PDF of $\alpha$ around its smallest allowed value.
We can also observe an impact on the PDF of the $\tilde f_{\rm NL}$ itself: e.g.\ if    $\cos \Delta$ is detected as being close to 1, then $\kappa_{\zeta}$ will be constrained by equation $(\zeta,\zeta\zeta)$ to $\kappa_{\zeta} \sim 1$. The detected value of $\alpha$ should be around $0.01$ (from current Planck constraints). Then, $(\zeta,\zeta S)$ gives $\tilde f_{\rm NL}^{\zeta,\zeta S} \sim \sqrt{\alpha} = 0.1$. In general, the error bars of $\tilde f_{\rm NL}$ from the bispectrum are much larger than 1, so this appears to be a significant improvement. But as we already said in section \ref{Planck joint analysis}, we cannot directly compare our results to those obtained in the Planck analysis, since, unlike Planck, we assume the model \eqref{fnl} as well as a flat prior on $\boldsymbol{\xi}$.

\item \label{point7} If we add to the previous case a detection of $\tilde f_{\rm NL}$,
or equivalently add the detection of $\cos\Delta$ to the case of point~\ref{point5}, any single $\tilde f_{\rm NL}$ suffices to detect the corresponding $\kappa_I$, while in point~\ref{point5} it had to be $\tilde f_{\rm NL}^{I,SS}$ or both the others. As in point~\ref{point6}, effects of contraction due to small detected $\cos\Delta$ can also occur. Again, once we have constraints on $\kappa_I$, the other relations allow improving constraints on $\beta_{\rm iso}$ and $\cos\Delta$.

\end{enumerate}
Next we will study models that predict a specific non-zero value of $\cos \Delta$. This is motivated by the curvaton scenario \cite{curvaton, curvaton1} that predicts a value equal to $\pm 1$. For more generality, we will also assume that there exist models predicting other values for the correlation. In these cases, we reduce the number of free parameters to three.

\begin{enumerate}
\setcounter{enumi}{7}
    \item \label{point8} Here, we assume that the correlation is fixed to a certain non-zero value, and that nothing is detected or fixed for the rest. The same mechanism already described in points \ref{point3} and \ref{point6} still holds: if we fix $\cos \Delta$ to a small enough value, $\kappa_I$ can be very large and still satisfy ($I,\zeta \zeta$), while at the same time providing strong constraints on $\alpha$ through $(I,\zeta S)$ or $(I,S S)$. Furthermore, like in the cases described in points \ref{point6} and \ref{point7} where the correlation parameter is detected, we have the possibility to improve the constraints on $\tilde f_{\rm NL}$, although the same caveats apply.
    
    \item \label{point9} If we have at least a detection of  $\tilde f_{\rm NL}^{I,\zeta\zeta}$, and independently of if we detect $\alpha$ in the power spectrum, the bispectrum allows to further constrain $\alpha$. Because in that case we can determine $\kappa_I$ from the ($I,\zeta\zeta$) equation, and use the other two equations to constrain $\alpha$. As in the previous case, that constraint will be better than with the power spectrum alone if $\cos \Delta$ is small.
    Furthermore, if we also detect another $\tilde f_{\rm NL}$, it can lead to a detection of $\alpha$.
\end{enumerate}
Other models, for example involving axion-like particles during inflation, predict uncorrelated adiabatic and isocurvature modes. For a review of the axion in cosmology, see \cite{axion}. Having $\cos\Delta=0$ reduces the six equations \eqref{fnl} to only two equations:
\begin{equation}
    \tilde f_{\rm NL}^{\zeta,SS}=\kappa_{\zeta}\alpha, \qquad \qquad \tilde f_{\rm NL}^{S,SS}=\kappa_{S}\alpha
\end{equation}

\begin{enumerate}
\setcounter{enumi}{9}
    \item \label{point10} Here, the two parameters $\kappa_I$ absorb all the constraints coming from the $\tilde f_{\rm NL}$. Using our model, the joint analysis cannot improve the constraints in the case of uncorrelated adiabatic and isocurvature modes if the $\kappa_I$ are free. This conclusion is independent of if we have a detection of $\alpha$ and $\tilde f_{\rm NL}$ or not.
    \item \label{point11} The only possibility to improve the constraints is to fix $\kappa_I$. Then we simply have that $\alpha=\tilde f_{\rm NL}^{I,SS} / \kappa_I$. So the more we fix $\kappa_I$ to a large value, the more the PDF of $\alpha$ contracts to zero. In the case of a non-detection of the $\tilde f_{\rm NL}$, we have for $\kappa \rightarrow \infty$: $\alpha \rightarrow 0$.
    \item \label{point12}Finally, if in addition to the previous point we have a detection of $\tilde f_{\rm NL}^{I,SS}$, then we have a detection of $\alpha$.
\end{enumerate}
Given all these theoretical results, we can now decide which cases will be the most interesting to study for each experiment. For Planck, we do not have any detection of isocurvature modes nor of $\tilde f_{\rm NL}$, neither in the power spectrum nor in the bispectrum. Hence, we already know that the joint analysis cannot help unless we fix $\kappa_I$ or $\cos \Delta$ as in points \ref{point3} and \ref{point8}. This explains the choices we made in the previous section. For future experiments like LiteBIRD and CMB-S4, we first have to determine if a detection is possible, and at what level, in the power spectrum and in the bispectrum, given the specifications of the instruments and the constraints from Planck.

\subsection{Future experiments}
\label{future}
In this section, we will determine for which region of parameter space the joint analysis will improve the constraints in the context of future experiments. We will then present the joint analysis results assuming a set of fiducial parameters in this region.

\subsubsection{Separate analyses of the power spectrum and the bispectrum} \label{Separate analysis}
We start by looking at the power spectrum alone.
To study forecasts for future experiments, we have to assume a true cosmology $(\boldsymbol{\theta}^0, \beta_{\rm iso}^0,\cos \Delta^0)$ compatible with the Planck data. Then using equation \eqref{clobs},  we determine the fiducial power spectra $\mathbf{ \tilde C_{\ell}}^{obs} = \mathbf{ \tilde C_{\ell} }({\boldsymbol{\theta}}^0, \beta_{\rm iso}^0,\cos \Delta^0)$. We will naturally set all the cosmological parameters $\boldsymbol{\theta}^0$ to their best estimated value given the Planck power spectrum with one non-vanishing isocurvature mode. We set the fiducial value of $\mathcal P_{SS}^{(1)}$ (from which $\beta_{\rm iso}$ is derived) to its $1\sigma$ upper value which is $\mathcal P_{SS}^{(1)|0}=4.4\times 10^{-11}$ for the CDM isocurvature mode, $\mathcal P_{SS}^{(1)|0}=1.7\times 10^{-10}$ for neutrino density and $\mathcal P_{SS}^{(1)|0}=1.1\times 10^{-10}$ for neutrino velocity. In the case of no detection, the $1\sigma$ (or any other) upper limit is computed using one tail. More explicitly, the one tail $1\sigma$ upper value means the largest value after excluding $32\%$ of the largest values.

\begin{figure}[t]
\centering
\includegraphics[scale=0.5]{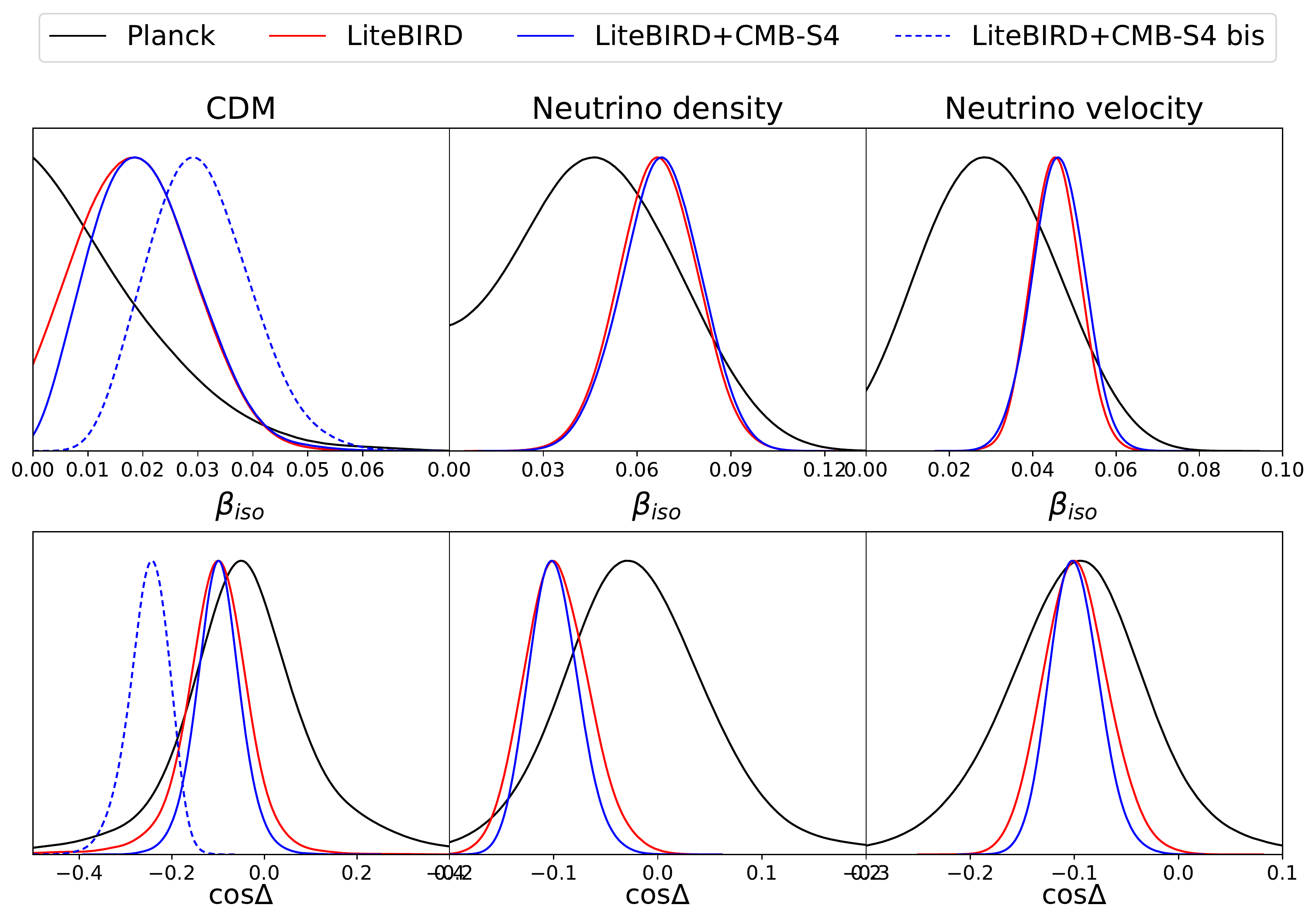}
\caption{ \label{Litecdm}  Marginalized PDF of $\beta_{\rm iso}$ (first row) and $\cos\Delta$ (second row) for Planck, LiteBIRD, and LiteBIRD+CMB-S4, for the three isocurvature modes CDM, neutrino density and neutrino velocity. These results are obtained from an analysis of the power spectrum alone. For the solid red and blue curves, we have chosen a fiducial value of $\beta_{\rm iso}$ at the $1\sigma$ upper limit of Planck and $\cos\Delta=-0.1$. The dashed blue curves show a more favourable case of fiducial values (for CDM only) of $\beta_{\rm iso}$ at the $1.5\sigma$ upper limit of Planck and $\cos\Delta^0=-0.25$ that will be used (and justified) for the joint analysis.}
\end{figure}

We choose $\cos \Delta^0=-0.1$, which is compatible with the Planck data for all three modes. To be in a more favourable case, we will push the fiducial value of $\beta_{\rm iso}$ to the upper limits of what is allowed by Planck, expressing this deviation from the Planck central value in terms of the number of $\sigma$ determined from the marginalized distribution of the parameter from Planck. However, to properly judge the (un)likeliness of the $\beta_{\rm iso}$ fiducial values that we choose, we should also take into account the chosen value of $\cos\Delta^0$, since, as we see for example in figure~\ref{planckcdmother} for the CDM isocurvature mode, these parameters are correlated. For example, a value of $\beta_{\rm iso}$ at the $1.5\sigma$ upper limit together with a small non-zero value of $\cos\Delta$ is actually likely at a level of $1\sigma$.\par
In figure \ref{Litecdm}, we show the marginalized constraints on $\beta_{\rm iso}$ in the first row and on $\cos \Delta$ in the second, for Planck, LiteBIRD and LiteBIRD+CMB-S4. LiteBIRD alone significantly improves all constraints compared to Planck, while adding CMB-S4 further improves the correlation parameter error bars by more than $20\%$. The CDM isocurvature mode has a low chance of being detected by a future experiment; we obtain at most a $2\sigma$ detection for LiteBIRD+CMB-S4 for the favourable configuration where the fiducial value of the isocurvature mode is at the $1\sigma$ upper limit of what is allowed by Planck. The dashed curves in the CDM plots of figure~\ref{Litecdm} correspond to an even more favourable configuration: the chosen fiducial parameters are $\mathcal P_{SS}^{(1)}=6.9\times 10^{-11}$, which is the $1.5\sigma$ upper value of Planck, and $\cos \Delta=-0.25$ that will be used in the joint analysis (this will be justified later). This configuration has a detection probability by LiteBIRD+CMB-S4 of at least $3\sigma$ for $\beta_{\rm iso}$ and more than $5\sigma$ for the correlation.
On the other hand, results are more promising for the neutrino density and velocity isocurvature modes, which would be detected at respectively $5\sigma$ and $7\sigma$ with the standard configuration described above. \par 

Next we consider the bispectrum alone. Unlike for the power spectrum, we will use here the sum of the Fisher matrices of all experiments: Planck+LiteBIRD+CMB-S4. Including the Planck likelihood for the power spectrum would bias our analysis, because we chose for the isocurvature parameters fiducial values different from those maximizing the Planck likelihood. Furthermore, adding the Planck power spectrum likelihood to the one for LiteBIRD and CMB-S4 does not improve the constraints on $\beta_{\rm iso}$ and $\cos\Delta$ significantly. However, for the bispectrum the situation is different. Adding the Planck Fisher matrix of the $\tilde f_{\rm NL}$ to the LiteBIRD and CMB-S4 Fisher matrices improves some $\tilde f_{\rm NL}$ constraints depending on the modes. Figures 1 and 2 of \cite{Akrami:2019izv} show that the constraints on the different modes are not equally improved by temperature and polarization measurements. For example, the neutrino density mode is mostly constrained (71\%) by temperature measurements alone. Hence Planck, which has a nearly optimal measurement of the temperature anisotropies, cannot be neglected. On the other hand, the polarization has a larger impact on the neutrino velocity mode, since temperature-only contributes here at a level of only 17\%. For these reasons, there is a benefit in considering jointly Planck, LiteBIRD and CMB-S4 for the bispectrum.

For this study, we need to fix each fiducial value $\tilde f_{\rm NL}^0$ such that they are compatible with the Planck results. Assuming our model, we determine each $\tilde f_{\rm NL}^0$ by using equations \eqref{fnl}. To do so, let us define the vector $\boldsymbol{\kappa}$ and the matrix $\mathcal M$: 
\begin{equation}\label{rewrite}
\boldsymbol{\kappa}=
\begin{pmatrix}
\kappa_{\zeta} & \\
\kappa_S &
\end{pmatrix},
\qquad \qquad
\mathcal M=
\begin{pmatrix}
\cos^2\Delta & \cos\Delta \sqrt{\alpha} & \alpha& 0&0&0&\\
0 & 0 & 0 & \cos^2\Delta & \cos\Delta \sqrt{\alpha} & \alpha
\end{pmatrix}
\end{equation} 
In order to choose the fiducial values $\kappa_I^0$ given $\beta_{\rm iso}^0$ and $\cos \Delta^0$, we want to determine the best estimated value of Planck given the PDF of equation~\eqref{probafnl}. Using equation~\eqref{rewrite}, we substitute $\mathbf {\tilde f_{\rm NL}}$ by $\mathbf{ \kappa}$ in \eqref{probafnl}:
\begin{equation}\label{probafnl2}
-2\ln P= \left( \mathcal{M}^T \boldsymbol{\kappa} -\mathbf { \tilde f_{\rm NL}^0} \right)^T  \mathbf{F}  \left( \mathcal{M}^T \boldsymbol{\kappa} - \mathbf {\tilde f_{\rm NL}^0} \right)
\end{equation}
The best estimated value, $ \boldsymbol{\hat \kappa}$, is the vector which maximizes the PDF \eqref{probafnl2}.
We find the following best estimated value and the covariance matrix of the parameters:
\begin{equation}\label{compatible}
    \boldsymbol{\hat \kappa} = \boldsymbol{\Sigma} \mathcal M \mathbf{F} \mathbf { \tilde f_{\rm NL}^0},
    \qquad \qquad
    \boldsymbol{\Sigma} = \left( \mathcal M \mathbf{F} \mathcal{M}^T \right)^{-1}
\end{equation}
For a fixed couple $\beta_{\rm iso}^0$ (and hence $\alpha^0$) and $\cos\Delta^0$, we now have to choose the $\kappa_I^0$ such that they give $\tilde f_{\rm NL}$ compatible with Planck measurements. Moreover, we define the signal-to-noise $N$ of $\tilde f_{\rm NL}$ as:
\begin{equation}
  \label{sigmaPC}
    N^{I,JK}(\kappa_I^0,\cos \Delta^0,\beta_{\rm iso}^0) = \frac{\left| \tilde f_{\rm NL}^{I,JK|0} (\kappa_I^0,\cos \Delta^0,\beta_{\rm iso}^0)\right| }{\sqrt{F^{-1}_{I,JK}}}
\end{equation}
where $\sqrt{F^{-1}_{I,JK}}$ means the marginalized errors on each $\tilde f_{\rm NL}^{I,JK}$ given a future experiment, which correspond to the square root of the diagonal entries of the inverse of the Fisher matrix. For LiteBIRD and CMB-S4, the values are given in table~\ref{errortable}.\par
\begin{figure}[t]
\centering
\includegraphics[scale=0.33]{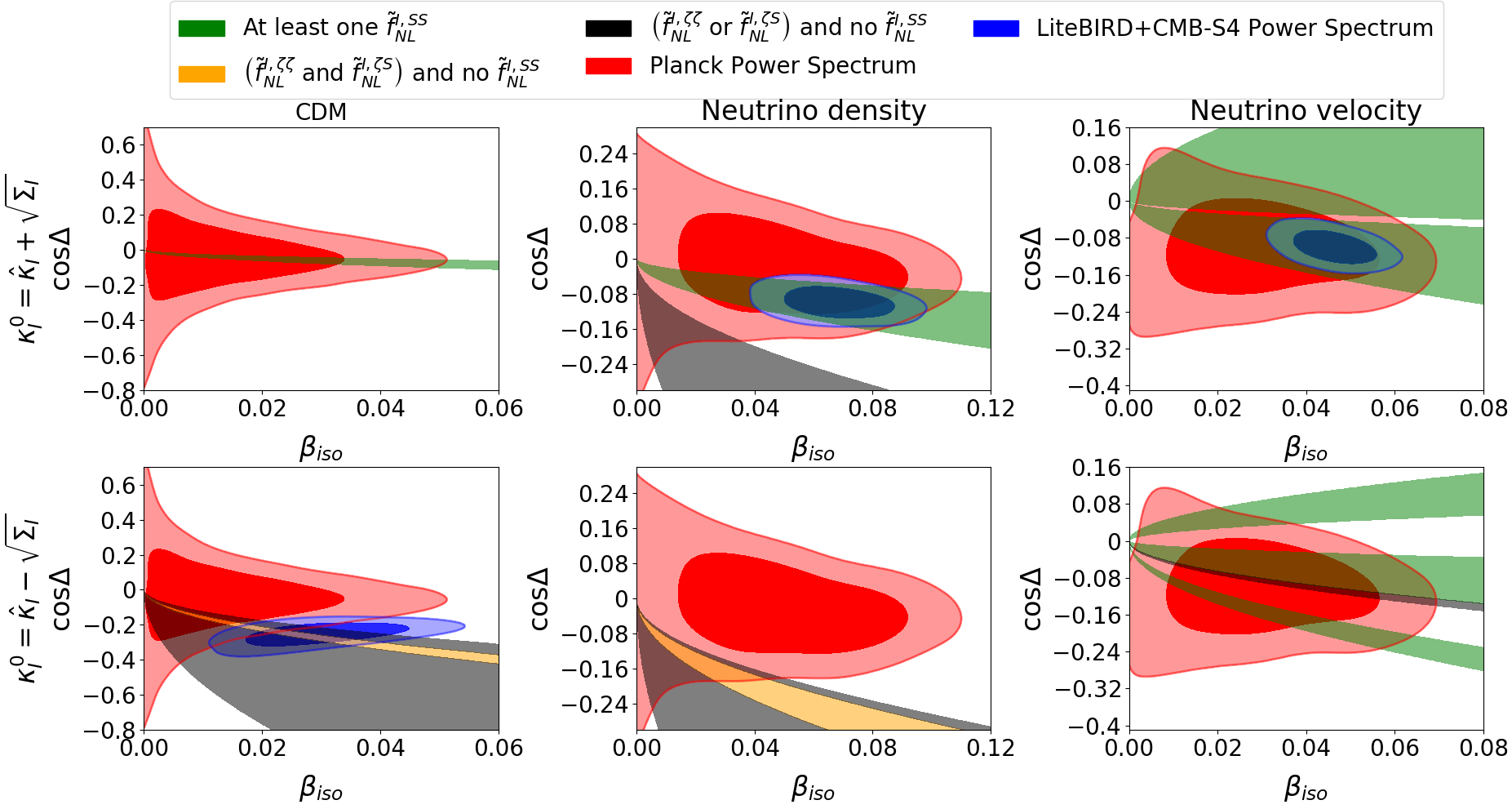}
\caption{ \label{nice} Constraints from the power spectrum alone for Planck, in red, and for LiteBIRD+CMB-S4, in blue (see the main text to understand the position of these blue contours). The green bands show the regions of the ($\beta_{\rm iso}, \cos\Delta$) space where one of the $\tilde f_{\rm NL}^{I,SS}$ will be detected by these future experiments given the indicated chosen value of $\kappa_I$. The region in orange indicates the set of fiducial parameters where none of the $\tilde f_{\rm NL}^{I,SS}$ are detected but where we detect the couple $(\tilde f_{\rm NL}^{I,\zeta \zeta},\tilde f_{\rm NL}^{I,\zeta S})$. Similarly, the region in black indicates the set of fiducial parameters where none of the $\tilde f_{\rm NL}^{I,SS}$ are detected but where we detect one (and one only) of the parameters $(\tilde f_{\rm NL}^{I,\zeta \zeta},\tilde f_{\rm NL}^{I,\zeta S})$. If we detect the isocurvature amplitude in the green or the orange region and if we do not detect the correlation parameter, then we are in the situation of point~\ref{point5} of section~\ref{Theoretical results} and the joint analysis will improve the constraints. If we detect the isocurvature amplitude in the green, orange or black regions and we also detect the correlation parameter, then we are in the situation of point~\ref{point7} and the joint analysis will also improve the constraints. The $\kappa_I^0$ are chosen at the $\pm 1\sigma$ value (first and second row, respectively). All the bands are calculated using the Planck+LiteBIRD+CMB-S4 constraints. }
\end{figure}

In the case where the isocurvature mode amplitude is detected in the power spectrum, the joint analysis of the power spectrum and the bispectrum will bring further constraints if we detect either the $\tilde f_{\rm NL}^{I,SS}$ or the couple $(\tilde f_{\rm NL}^{I,\zeta\zeta},\tilde f_{\rm NL}^{I,\zeta S})$ for a fixed $I$, which gives us 4 possibilities in total (see point~\ref{point5} of section~\ref{Theoretical results}). If, in addition, we also detect the correlation, only one $\tilde f_{\rm NL}$ needs to be detected in order to improve the constraints (see point~\ref{point7} of section~\ref{Theoretical results}). In figure~\ref{nice}, we give the $(\beta_{\rm iso},\cos\Delta)$ constraints from Planck in red. In order to determine for which region of parameter space the constraints would be improved by the joint analysis, we calculate for each couple ($\beta_{\rm iso},\cos\Delta$), the best estimation $\boldsymbol{\hat\kappa}$ and the error $\Sigma$ using equation~\eqref{compatible}. Then, using equation~\eqref{sigmaPC}, we calculate the signal-to-noise $N^{I,JK}(\hat\kappa_I\pm\sqrt{\Sigma_I},\beta_{\rm iso},\cos\Delta)$. The green bands correspond to the region of the parameter space where at least one $N^{I,SS}$ is larger than $4$. Similarly, the orange bands correspond to $(N^{I,\zeta \zeta}>4$ and $N^{I,\zeta S}>4)$ and $N^{I,SS}<4$ and the black bands correspond to ($N^{I,\zeta \zeta}>4$ or $N^{I,\zeta S}>4$) and $N^{I,SS}<4$. The signal-to-noise coefficients have been calculated using the Planck+LiteBIRD+CMB-S4 Fisher matrix. \par
Given point~\ref{point5} of section~\ref{Theoretical results}, if we detect the amplitude of the isocurvature mode but not the correlation using the power spectrum alone, the joint analysis would improve constraints for the ensemble of the fiducial values represented in green and orange. Given point~\ref{point7} of section~\ref{Theoretical results}, if we detect both the amplitude and the correlation with the power spectrum alone, the joint analysis would improve the constraints for the ensemble of fiducial parameters represented in green, orange and black. The first (second) row corresponds to a fiducial value of $\kappa$ such that the $\tilde f_\mathrm{NL}$ are at the $1\sigma$ upper (lower) value from their maximum probability for the Planck data. The blue contours correspond to the solid blue curves in figure~\ref{Litecdm} for neutrino density and neutrino velocity and to the dashed blue curve for CDM.\par

Let us choose for each mode a couple $(\beta_{\rm iso}^0, \cos \Delta^0)$ in one of the bands that will give at least a detection of the amplitude in the power spectrum. For CDM, the $1\sigma$ upper value of $\kappa_I^0$ leaves us a very thin green band close to $\cos\Delta=0$. We could choose our fiducial parameters in this band, but in order to detect the amplitude in the power spectrum, we need a $2\sigma$ Planck compatible value of $\beta_{\rm iso}$. Instead we choose the $1\sigma$ lower value of $\kappa_I^0$, but this still requires fiducial values at the edge of the $2\sigma$ Planck contour in order to have a detection of $\beta_{\rm iso}$, $\cos\Delta$ and satisfy point~\ref{point7}. Our choice is arbitrary since none of the two possibilities is statistically more likely. The configuration we choose gives the marginalised PDF of the dashed blue curves in the CDM plots of figure~\ref{Litecdm}.\par 
For neutrino density and velocity, the total parameter space, in which the fiducial values can be chosen in order to have better constraints with the joint analysis, is larger than for the CDM case. In particular for the neutrino velocity mode more than half of the $1\sigma$ Planck contour is covered by the green band, as can be seen in the top right of figure~\ref{nice}. Instead of having to consider a more favourable case as for CDM, for the neutrino modes we can safely keep the fiducial values used for the solid blue curves in the neutrino plots of figure~\ref{Litecdm} and take the $1\sigma$ upper value for $\kappa_I$. 
The power spectrum analysis for LiteBIRD+CMB-S4 given all these final fiducial values gives us the blue contours in figure~\ref{nice}. We have chosen not to put any blue contours in the subplots with the values of $\kappa$ that we do not use for our subsequent analysis.\par

\subsubsection{Joint analysis} \label{Joint analysis}
In section \ref{Separate analysis}, we have determined the fiducial values of the isocurvature parameters for which the joint analysis will provide a clear improvement compared to an analysis with the power spectrum alone. We choose the fiducial values for the isocurvature mode power spectrum amplitude $\mathcal P_{SS}^{(1)|0}=6.9\times 10^{-11}$ for CDM, $\mathcal P_{SS}^{(1)|0}=1.7\times 10^{-10}$ for neutrino density and $\mathcal P_{SS}^{(1)|0}=1.1\times 10^{-10}$ for neutrino velocity, which are compatible with the Planck results while being significantly detectable by LiteBIRD, thus avoiding the parameters $\kappa_I$ to be unbounded which would lead to results that are difficult to interpret. Furthermore, we have shown that the fiducial value of the correlation $\cos\Delta$ has a strong impact on the $\tilde f_{\rm NL}$ detection. In order to see the effect of the bispectrum constraints, we have chosen the correlation and the $\kappa_{I}$ such that they verify point~\ref{point7} for CDM and point~\ref{point5} for neutrino density and velocity. Fiducial parameters are for CDM $\mathcal P_{\zeta S}^{(1)|0}=-1.0\times 10^{-10}$, for neutrino density  $\mathcal P_{\zeta S}^{(1)|0}=-6.3\times 10^{-11}$ and for neutrino velocity $\mathcal P_{\zeta S}^{(1)|0}=-5.1\times 10^{-11}$. \par

\begin{figure}[t]
\centering
\includegraphics[scale=0.40]{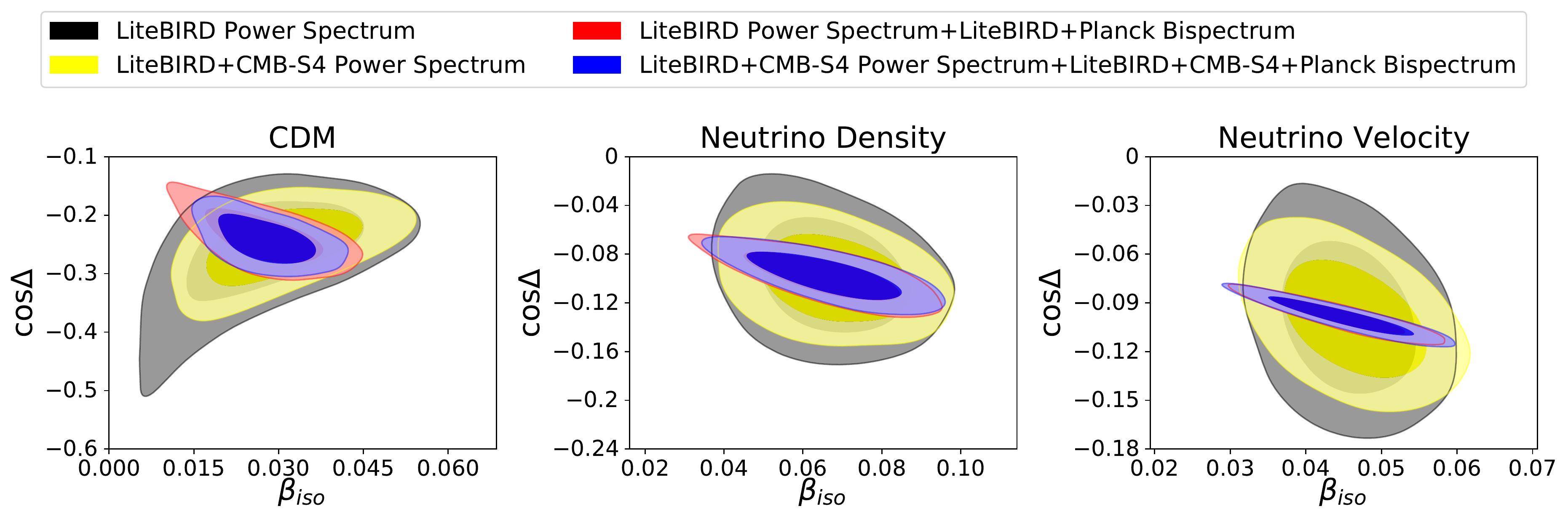}
\caption{Constraints in the $(\beta_{\rm iso},\cos \Delta)$ space for the case of CDM (left), neutrino density (centre), and neutrino velocity isocurvature (right). The different colours show constraints from the power spectrum alone and from the joint analysis of the power spectrum and the bispectrum, of LiteBIRD with and without CMB-S4, as indicated in the legend. \label{Litecdmresult}}
\end{figure}
 \begin{table}[t]
 \centering
 \begin{tabular}{ | l l |c c c c c|}
\hline 
&&$\Omega_b h^2$ & $\Omega_c h^2$& $\tau$& $100\theta_{MC}$ & $\log 10^{10} A_s$\\
 \hline 
CDM&PS&5.1e-05&0.00044&0.0023&0.00013&0.0058\\
 \hline 
&PS+B&5.1e-05&0.00044&0.0022&0.00013&0.0053\\
 \hline 
&improvement&No&No&No&No&9\%\\
 \hline 
ND&PS&4.5e-05&0.00038&0.0023&0.00015&0.0063\\
 \hline 
&PS+B&4.6e-05&0.00037&0.0023&0.00012&0.0051\\
 \hline 
&improvement&No&No&No&20\%&19\%\\
 \hline 
NV&PS&4.5e-05&0.00039&0.0025&0.00015&0.0082\\
 \hline 
&PS+B&4.7e-05&0.00038&0.0025&0.00011&0.0057\\
 \hline 
&improvement&No&No&No&30\%&31\%\\
 \hline 
\hline 
&&$n_s$ & $\beta_{\rm iso}$ & $\cos \Delta$ & $\kappa_{\zeta}$ & $\kappa_{S}$\\
 \hline 
CDM&PS&0.0021&0.009&0.045&&\\
 \hline 
&PS+B&0.0022&0.006&0.028&133&1767\\
 \hline 
&improvement&No&35\%&38\%&&\\
 \hline 
ND&PS&0.0017&0.012&0.024&&\\
 \hline 
&PS+B&0.0015&0.013&0.013&428&3567\\
 \hline 
&improvement&10\%&No&45\%&&\\
 \hline 
NV&PS&0.0016&0.006&0.024&&\\
 \hline 
&PS+B&0.0016&0.006&0.008&897&549\\
 \hline 
&improvement&No&No&67\%&&\\
 \hline 
\end{tabular}
\caption{Marginalized $1\sigma$ uncertainties of the six cosmological parameters and the four parameters of our model obtained for each isocurvature mode (ND/NV being neutrino density and neutrino velocity, respectively) from the LiteBIRD+CMB-S4 power spectrum (PS) likelihood and from the LiteBIRD+CMB-S4+Planck bispectrum likelihood for the joint analysis (PS+B). The third line of each isocurvature mode shows the percentage of improvement of the error bars for the joint analysis compared to the analysis of the power spectrum alone. If the absolute value of the improvement is smaller than $5\%$, we consider it as being not significant and then write simply 'No'. \label{errtable}}
 \end{table}

The results for the $\beta_{\rm iso}$ and $\cos\Delta$ constraints are shown in figure \ref{Litecdmresult} for each isocurvature mode, both for the analysis of the power spectrum alone and for the joint analysis, and both excluding and including the contribution of CMB-S4. We show results in the $(\beta_{\rm iso}, \cos\Delta)$ space, because we expect the bispectrum to bring further constraints in this parameter space. As we have seen in figure \ref{Litecdm}, the addition of CMB-S4 to LiteBIRD does not significantly improve the marginalized $\beta_{\rm iso}$ constraints in a power-spectrum-only analysis, while there is some improvement for $\cos \Delta$. As always, the marginalized distributions just contain partial information; here in figure~\ref{Litecdmresult} we can see the improvement of the 2D contours (yellow versus gray). We also see a small improvement in the joint analysis results when adding CMB-S4 (blue versus red). The most important contribution of CMB-S4, in our analysis, is to increase the detection of some $\tilde f_{\rm NL}$ and thus to increase the size of the bands in parameter space where the joint analysis is useful, see figure~\ref{nice}. The quantitative results of the rest of this section, for example the error bars summarized in table~\ref{errtable}, are for LiteBIRD+CMB-S4. \par

The CDM isocurvature mode constraints are improved significantly by the joint analysis of the power spectra and bispectra.
We detect in this case $\tilde f_{\rm NL}^{S,\zeta\zeta}$ $(5.7\sigma)$ and $\tilde f_{\rm NL}^{S,\zeta S}$ ($4.0\sigma$). Thanks to the relation $(S,\zeta\zeta)$ of~\eqref{fnl}, we detect $\kappa_S$ and obtain: $-6243^{+3564}_{-5830}$ (99\% confidence level), while the fiducial value is $-5788$. The relation $(S,SS)$ then improves the uncertainty of $\beta_{\rm iso}$ to $0.006$. Thus, in table~\ref{errtable} we see that adding the bispectrum improves the detection of $\beta_{\rm iso}$ by $35\%$. Moreover, very small values of $\cos\Delta$ are suppressed, which improves the error bars of the correlation by $38\%$. 
\par
For the neutrino density isocurvature mode, we detect $\tilde f_{\rm NL}^{S,SS}$ at a level of $7\sigma$. This allows a detection of $\kappa_S$ with a measured value of $13362^{+11539}_{-7056}$ (99\% CL) for a fiducial value of $12611$. The constraints on $\tilde f_{\rm NL}^{S,\zeta\zeta}$ and $\tilde f_{\rm NL}^{S,\zeta S}$ improve the error bars of the correlation $\cos\Delta$ by $45\%$. However, there is no improvement of $\beta_{\rm iso}$. \par

For the neutrino velocity isocurvature mode, both $\tilde f_{\rm NL}^{\zeta,\zeta S}$ and $\tilde f_{\rm NL}^{\zeta,S S}$ are detected, at a level of $11\sigma$ and $7\sigma$, respectively. Thanks to relation $(\zeta,SS)$, $\kappa_{\zeta}$ is detected and we find $4836^{+3073}_{-1741}$ (99\% CL) for a fiducial value of $4653$. Actually we also detect $\tilde f_{\rm NL}^{S,SS}$ at the level of $4\sigma$, but this has only a weak influence on the error bars of $\beta_{\rm iso}$ and $\cos\Delta$, although it gives a detection of $\kappa_{S}$ with a measured value of $2455^{+1849}_{-1133}$ (99\% CL) for a chosen fiducial value of $2346$. Relation $(\zeta,\zeta S)$ provides a detection of the correlation $\cos\Delta$ at the level of $12\sigma$. The error bars of the correlation parameter are also constrained significantly thanks to the $(\zeta,\zeta\zeta)$ relation. The final uncertainty on the correlation parameter will shrink by $67\%$ in this configuration thanks to the joint analysis. As in the case of neutrino density, the parameter $\beta_{\rm iso}$ is not affected since all $\tilde f_{\rm NL}$ constraints are absorbed by $\kappa_I$ and $\cos\Delta$. \par

\begin{figure}[t]
\centering
\includegraphics[scale=1]{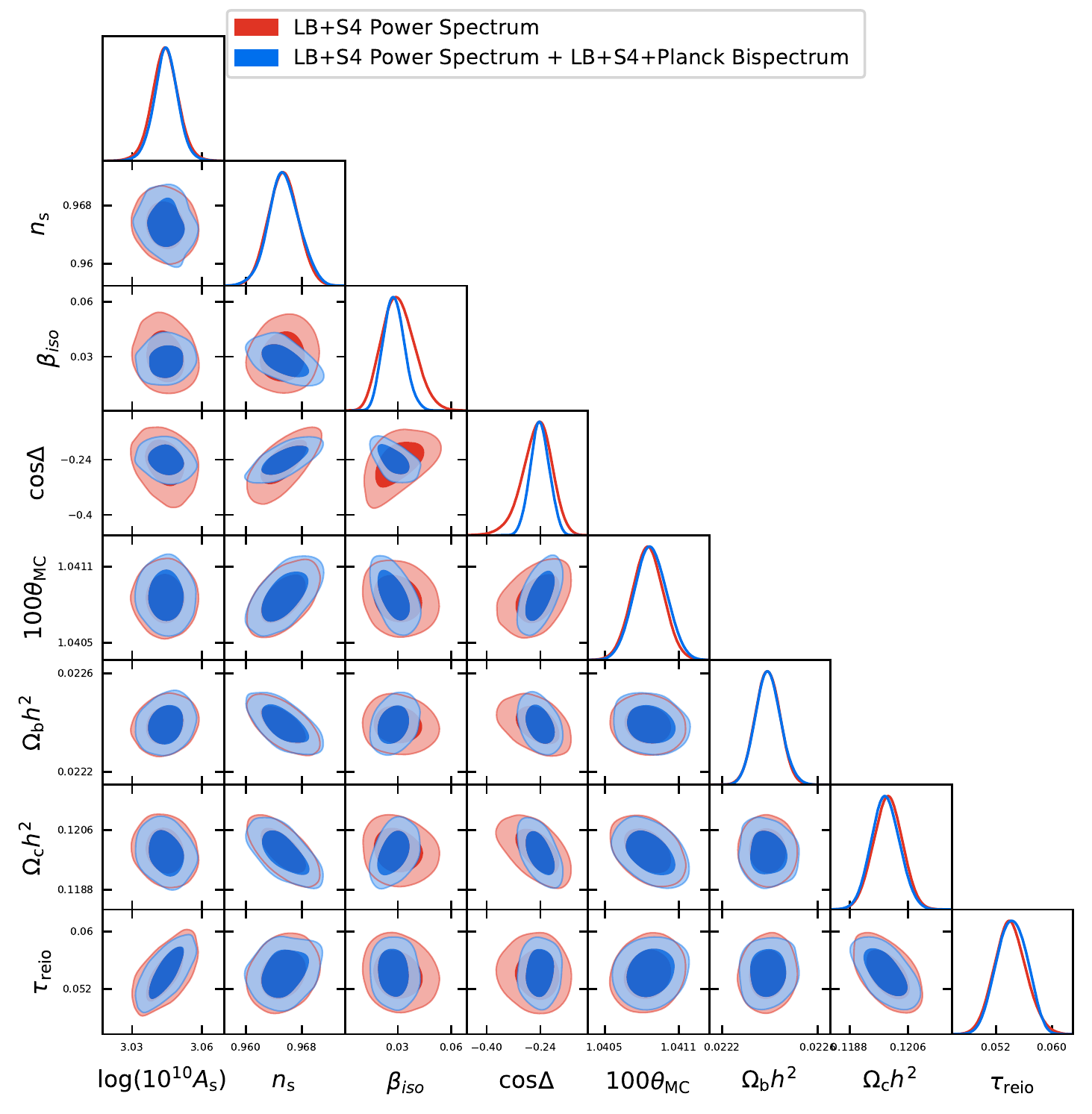}
\caption{Two dimensional 68\% and 95\% contours assuming $\Lambda$CDM plus a CDM isocurvature mode for LiteBIRD+CMB-S4 (+Planck for the bispectrum). The red contours show the results from the power spectrum alone and the blue contours the results from the joint analysis.\label{Future_cos_cdm} }
\end{figure}
\begin{figure}[t]
\centering
\includegraphics[scale=1]{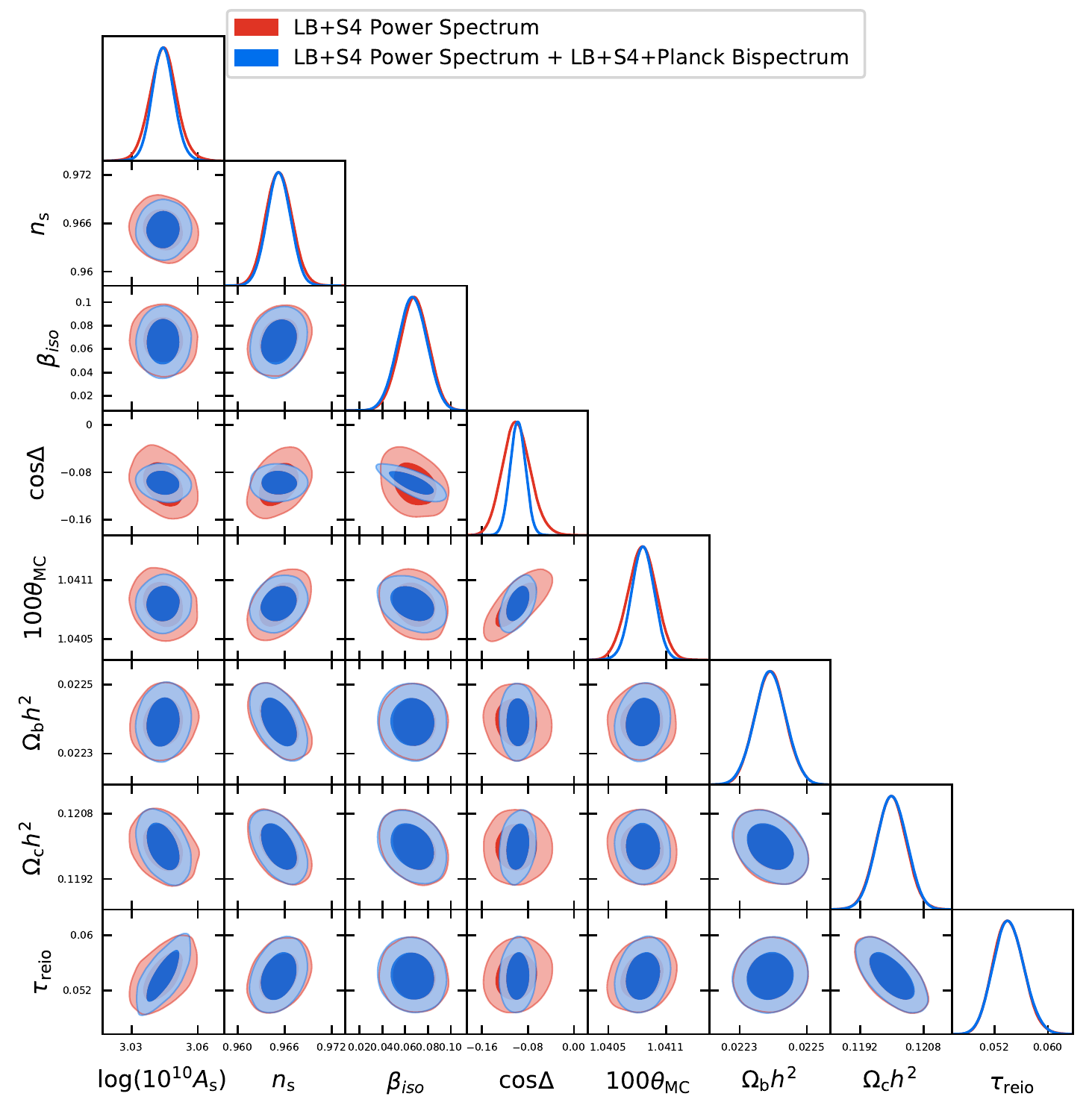}
\caption{Same as figure~\ref{Future_cos_cdm} but for the neutrino density isocurvature mode. \label{Future_cos_nu} }
\end{figure}
\begin{figure}[t]
\centering
\includegraphics[scale=1]{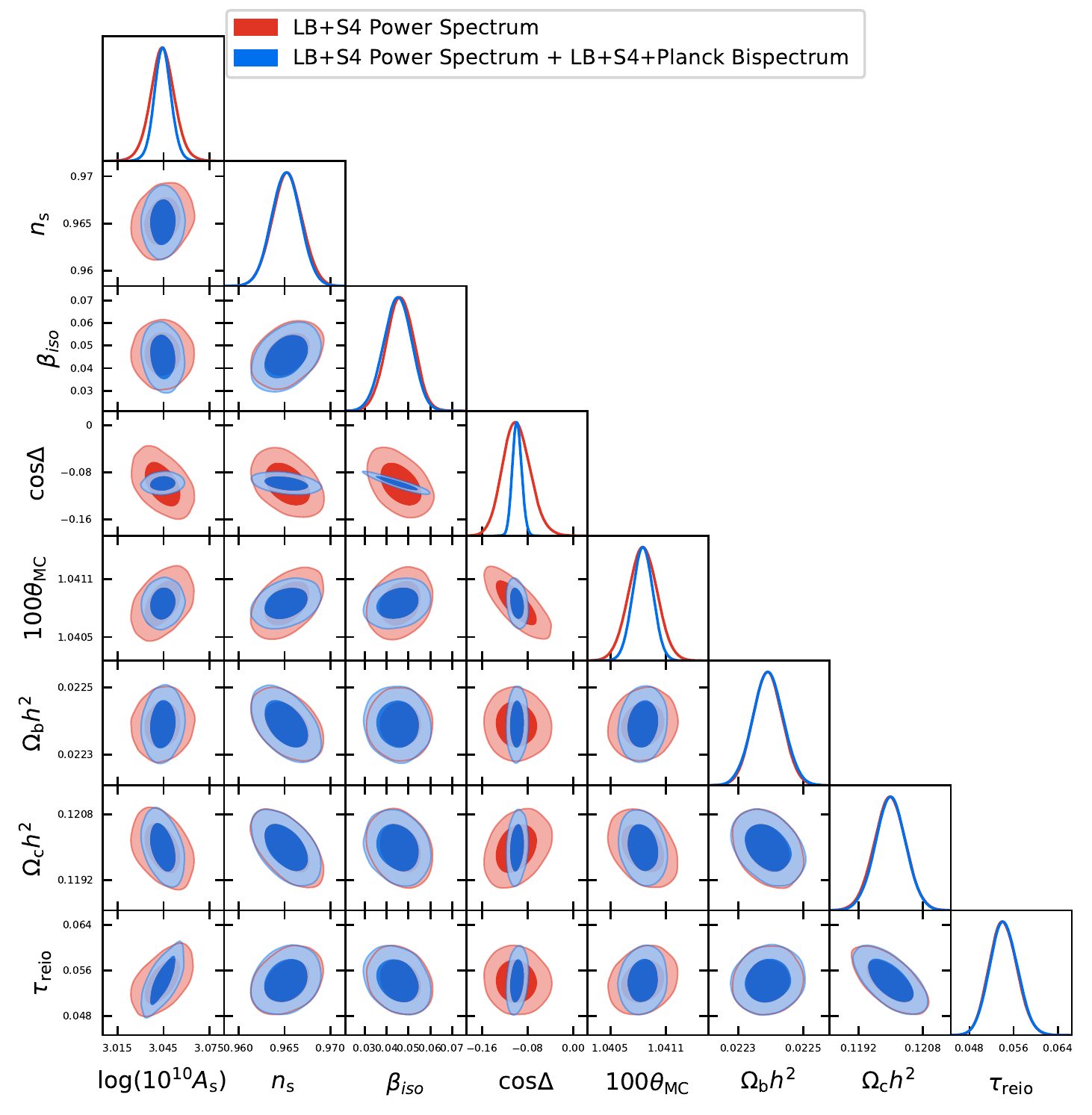}
\caption{Same as figure~\ref{Future_cos_cdm} but for the neutrino velocity isocurvature mode. \label{Future_cos_vnu}}
\end{figure}

Figures~\ref{Future_cos_cdm}, \ref{Future_cos_nu} and \ref{Future_cos_vnu} show the 2D contours for all pairs of the parameters of the model (excluding the $\kappa$ parameters), and provide an estimation of the correlation between the parameters.
For all isocurvature modes we observe an anti-correlation between $\beta_{\rm iso}$ and $\cos\Delta$ (except for the case of CDM power spectrum only because of the lack of a detection of $\beta_{\rm iso}$). It can be understood as follows: both $\beta_{\rm iso}$ and the correlation parameter lead to an increase of the power spectrum, so one parameter can be compensated by the other and lead to a similar amplitude of the power spectrum.

The improvement of the constraints on the correlation parameter $\cos\Delta$ coming from the bispectrum also induces improvements on the constraints of the cosmological parameters that are correlated with $\cos \Delta$, as can be seen in table~\ref{errtable}. 
The cosmological parameters that are correlated most with $\cos\Delta$ for all isocurvature modes are $A_s$, $n_s$, and $\theta_{MC}$. We observe for the neutrino modes in figures~\ref{Future_cos_nu} and \ref{Future_cos_vnu} an anti-correlation between $\cos\Delta$ and $A_s$, which is always suppressed by the joint analysis. The marginalized error of $A_s$ is improved by $19\%$ and $31\%$ for neutrino density and neutrino velocity, respectively. We see in those figures a reduction of the  $A_s$ uncertainty independently of $\cos \Delta$. This means that the bispectrum constrains $A_s$ directly. The effect of the bispectrum is very weak in the case of the CDM isocurvature mode in figure~\ref{Future_cos_cdm}, only $9\%$ of improvement, probably because this mode is only detected at 2$\sigma$.\par

The constraints on $n_s$ mostly come from the relative amplitude of the power spectrum between small $\ell$ and large $\ell$. The CDM isocurvature mode contributes most at low $\ell$. As in figure 2 of \cite{Langlois:2012tm}, increasing $\cos\Delta$ will increase the low-$\ell$ part of the total power spectrum, which corresponds to a small $n_s$. It can then be compensated by a larger $n_s$ which means that the parameters are correlated. For neutrino density, the relative amplitude between low-$\ell$ and the second peak is almost unity, while for the adiabatic mode, the second peak is roughly two times higher. Thus, increasing $\cos\Delta$ will decrease the ratio between the low-$\ell$ and the second peak amplitudes, which corresponds to a smaller $n_s$. This leads to a correlation between $\cos\Delta$ and $n_s$ as with the CDM isocurvature mode. On the contrary, the neutrino velocity isocurvature mode has a larger contribution to the second peak compared to the low-$\ell$ part (see once more figure~2 of \cite{Langlois:2012tm}). Hence we find here an anti-correlation between $\cos\Delta$ and $n_s$. The joint analysis breaks this correlation only for neutrino density and then improves the marginalized error bar of $n_s$ by 10\% in that case. No significant improvement is observed for the other modes.\par

The constraints on $\theta_{MC}$ come from the positions of the peaks in the power spectrum. As we can see in figure~2 of \cite{Langlois:2012tm}, all isocurvature-adiabatic cross power spectra are phase shifted compared to the pure adiabatic mode. Thus increasing $\cos\Delta$ will automatically shift the position of the peaks and hence directly affect the estimation of $\theta_{MC}$. As can be seen in figures~\ref{Future_cos_cdm} and \ref{Future_cos_nu}, for the CDM and neutrino density isocurvature modes, there is a positive correlation between the $\cos\Delta$ and $\theta_{MC}$ parameters, because increasing $\cos\Delta$ shifts the position of the peaks to higher $\ell$ since the density isocurvature modes have their first peaks on the right of the adiabatic first peak. On the contrary, the neutrino velocity isocurvature mode, which is roughly the derivative of the neutrino density mode and hence is in counterphase with the latter, has its peak on the left of the adiabatic one. This gives an anti-correlation between the $\cos\Delta$ and $\theta_{MC}$ parameters. The joint analysis, by improving the $\cos\Delta$ constraint, is then able to improve the $\theta_{MC}$ error bar by 20\% and 30\% for the neutrino density and velocity isocurvature modes, respectively. There is only a small improvement for the CDM isocurvature mode since the correlation between $\cos\Delta$ and $\theta_{MC}$ is weak. \par

\subsubsection{Excluding the model with future experiments}
\begin{figure}
    \centering
    \includegraphics[scale=1]{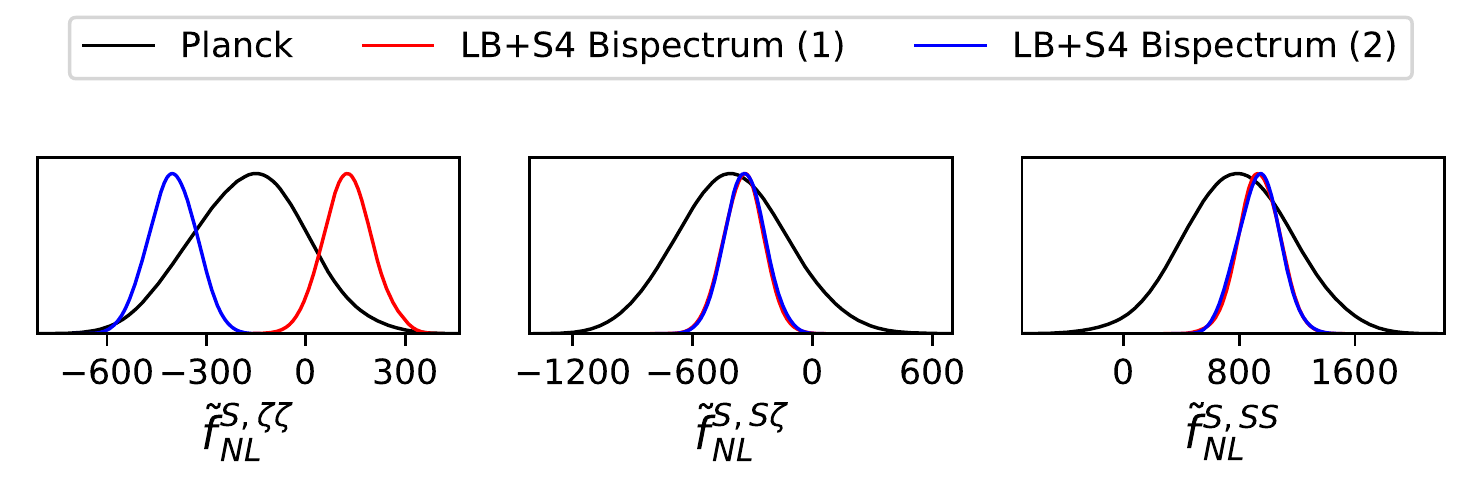}
    \caption{Marginalized constraints for $\tilde f_{\rm NL}^{S,IJ}$ for the case of neutrino density isocurvature. In black the model-independent (bispectrum-only) Planck results. In red, the bispectrum-only LiteBIRD+CMB-S4 constraints, assuming fiducial values computed with \eqref{fnl} using the same $(\beta_\mathrm{iso}^0, \cos\Delta^0, \kappa_I^0)$ as in section \ref{Joint analysis}. The blue curves are similar to the red ones but with a different fiducial value for $\tilde f_{\rm NL}^{S,\zeta\zeta}$.}
    \label{ffnu}
\end{figure}
\begin{figure}
    \centering
    \includegraphics[scale=0.95]{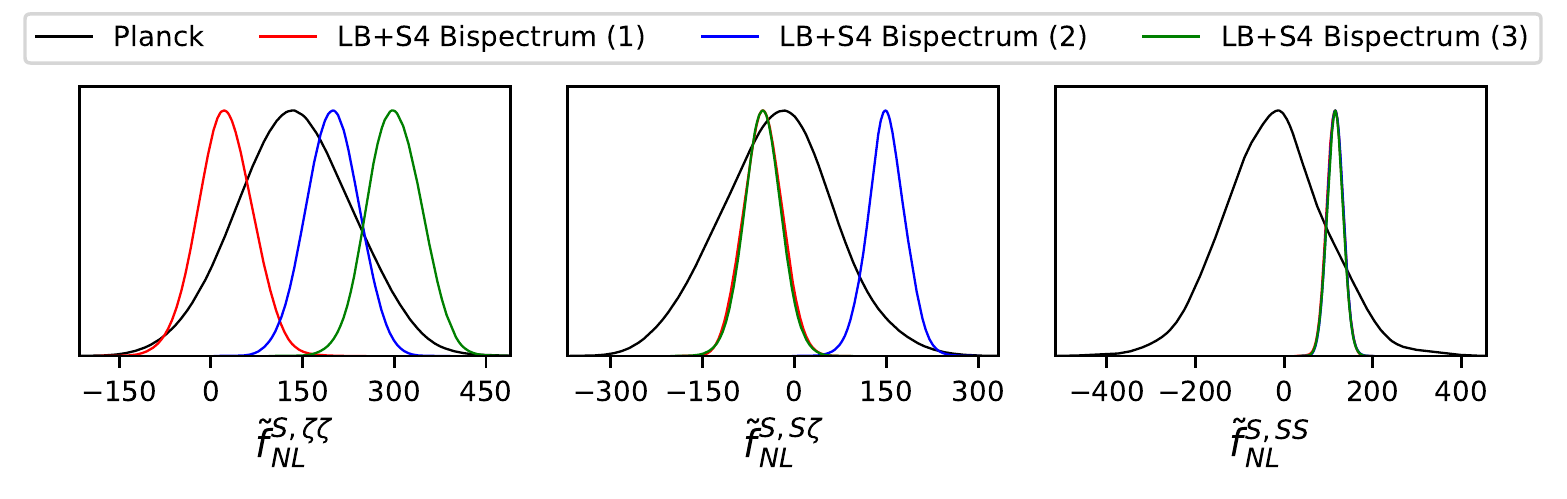}
    \caption{Marginalized constraints for $\tilde f_{\rm NL}^{S,IJ}$ for the case of neutrino velocity isocurvature. Black and red curves are the same as in figure~\ref{ffnu} but for neutrino velocity. The blue and green curves are similar to the red ones but with different fiducial values for $\tilde f_{\rm NL}^{S,\zeta\zeta}$ (blue and green) and $\tilde f_{\rm NL}^{S,\zeta S}$ (blue only).}
    \label{ffvnu}
\end{figure}

Up to this point, we have investigated the improvements a joint analysis of the power spectrum and the bispectrum can provide regarding the constraints (or detection) of the isocurvature parameters. To make a joint analysis possible, we assumed the quite general class of inflation models described in section \ref{two-field}. However, it is also interesting to see if it would be possible to rule out this class of inflation models with LiteBIRD and CMB-S4. In fact this turns out to be the case, if the values detected by LiteBIRD/CMB-S4, while being compatible with Planck, do not satisfy the relations discussed in section~\ref{two-field}. \par

Among those relations, equation \eqref{fnllink} implies that $\tilde f_{\rm NL}^{I,\zeta \zeta}$ and $\tilde f_{\rm NL}^{I,SS}$ must share the same sign, and this can be tested using just a bispectrum-only analysis. In figure \ref{ffnu}, we show Planck's model-independent (bispectrum-only) estimation of the $\tilde{f}_\mathrm{NL}^{S,IJ}$ for the case of neutrino density in black. The red and blue curves are two LiteBIRD+CMB-S4 bispectrum constraints that differ in their choice of fiducial value for $\tilde f_{\rm NL}^{S,\zeta\zeta}$: the red curve corresponds to the choices made in section~\ref{Joint analysis}, while for the case in blue we have changed the fiducial value of $\tilde f_{\rm NL}^{S,\zeta\zeta}$ in order to get a detection at a negative value. As in this blue case $\tilde f_{\rm NL}^{I,\zeta \zeta}$ and $\tilde f_{\rm NL}^{I,SS}$ have different signs at a high level of confidence, such a detection would rule out the class of models~\eqref{ourmodel}.

The test above required only an analysis of the bispectrum. Next we investigate if adding information from the power spectrum can rule out the model in cases where the bispectrum alone would not suffice. In figure \ref{ffvnu}, which is similar to figure \ref{ffnu} but for neutrino velocity, we show again the Planck model-independent bispectrum-only constraints in black and, this time, three cases of LiteBIRD+CMB-S4 constraints with different Planck-compatible fiducial values, in red, blue and green. The red curves show the same choice of fiducial parameters as in section~\ref{Joint analysis}, and hence they are by construction compatible with the model. The blue curves show a set of fiducial values that would invalidate the model thanks to combining information from the power spectrum and the bispectrum. Indeed, we have here a detection of all $\tilde f_{\rm NL}^{S,IJ}$ at a positive value. However, if we also detect a negative correlation $\cos\Delta$ in the power spectrum like in section \ref{Separate analysis}, the model would be ruled out, since another prediction of the model is that if $\tilde f_{\rm NL}^{I,\zeta \zeta}$ and $\tilde f_{\rm NL}^{I,SS}$ are both positive, then $\tilde f_{\rm NL}^{I,\zeta S}$ must have the same sign as the correlation $\cos \Delta$.\par

The third case in figure \ref{ffvnu}, in green, corresponds also to a set of fiducial values that would rule out the model thanks to combining information from the power spectrum and the bispectrum. In this case, $\tilde f_{\rm NL}^{S,SS}$ is detected at more than $6\sigma$: $114\pm 18$. If we assume that we detect the relative amplitude $\beta_\mathrm{iso}$ (and hence $\alpha$) and the correlation $\cos\Delta$ of the neutrino velocity isocurvature mode thanks to the power spectrum at the same values as in section \ref{Joint analysis}, then equation $(S,SS)$ of \eqref{fnl} would give us $\kappa_S = 2425\pm 526$. That, in its turn, would lead to the following prediction from equation $(S,\zeta\zeta)$: $\tilde f_{\rm NL}^{S,\zeta\zeta}=23 \pm 13$. However, as we see in figure \ref{ffvnu}, this is excluded at more than $5\sigma$ with the bispectrum analysis. \par

These three examples show that LiteBIRD+CMB-S4 will potentially be able to exclude the class of models~\eqref{ourmodel} that we have assumed for the joint analysis, depending of course on what values for the different parameters will finally be observed. In the first example, the bispectrum-only analysis is enough to exclude the model. In the other two examples, we need to add information from the power spectrum. Thus, combining information from the power spectrum and the bispectrum can allow us to check if the model is consistent with the data.

\section{Conclusion }\label{concl}
The presence of isocurvature modes (in addition to the dominant adiabatic mode) in the CMB would be a direct proof that the cosmological perturbations are produced by at least two primordial degrees of freedom, which in the context of the inflationary paradigm would mean multi-field inflation. Hence this would rule out single-field inflation, which for the moment is still consistent with all observations. Given the matter content of the universe we have three possible isocurvature modes: CDM density, neutrino density, and neutrino velocity (a fourth possibility, a baryon density isocurvature mode, is observationally indistinguishable from the CDM mode in the CMB, and hence will not be considered separately here). The Planck power spectrum analysis did not find any sign of these isocurvature modes, and put tight constraints on their amplitudes \cite{Akrami:2018odb}. Similarly the Planck bispectrum analysis did not detect any isocurvature non-Gaussianity (nor any other type of primordial non-Gaussianity in fact) \cite{Akrami:2019izv}.\par

In this paper we have performed a joint analysis of the power spectrum and the bispectrum in order to improve the isocurvature constraints using the Planck data, and we have made forecasts for the future satellite LiteBIRD and the future ground-based CMB-S4 experiments. To do so, we need to assume a model that allows us to express both the power spectrum observables and the bispectrum observables in terms of a set of common model parameters. In our choice of model we had to make a compromise between, on the one hand, keeping the model as general as possible so that our analysis applies to as broad a class of inflation models as possible, and, on the other hand, keeping the number of additional free parameters limited, otherwise we do not get any meaningful constraints. This compromise led us in the first place to consider only two-field inflation models, with a single isocurvature mode (which can be any of the three mentioned above) in addition to the adiabatic mode. Secondly, we assumed that one of the fields dominates both the linear isocurvature mode and the second-order (non-Gaussian) parts of the adiabatic and the isocurvature mode, the other field only contributing to the linear adiabatic mode (see \eqref{ourmodel}). For the rest, however, this model is completely general. It is the same model as considered in the last section of \cite{Langlois:2012tm} and has five free parameters, one of which is fixed by the adiabatic amplitude of the power spectrum. Hence our model has four extra parameters compared to the standard $\Lambda$CDM cosmology, which can be viewed as the isocurvature amplitude of the power spectrum $\beta_{\rm iso}$, the linear correlation between the adiabatic and the isocurvature mode $\cos\Delta$, and the adiabatic and isocurvature bispectrum amplitudes $\kappa_\zeta$ and $\kappa_S$. The power spectrum only depends on the first two. As was explained in \cite{Langlois:2012tm}, in such a configuration there are six different $\tilde f_\mathrm{NL}^{I,JK}$ parameters (with $I,J,K=\zeta,S$ and symmetric under interchange of $J$ and $K$, where $\zeta$ indicates the adiabatic mode and $S$ the isocurvature mode) that can be extracted from the bispectrum, although the relations imposed by the model mean that only three of them are independent.\par

First we applied our methodology to the Planck data. We built a joint power spectrum and $\tilde f_\mathrm{NL}$ likelihood, which is simply the product of the two likelihoods as we argue that they can be considered to be statistically independent. We used the Planck 2018 likelihood for the power spectrum. In addition to the cosmological parameters, we have estimated all the nuisance parameters including those of the foregrounds. As a full bispectrum likelihood cannot be calculated, we consider a much simpler $\tilde f_\mathrm{NL}$ likelihood based on the Fisher matrix.
This is nonetheless a nearly optimal procedure, because it is directly related to the fact that the $\tilde f_\mathrm{NL}$ estimator \eqref{ideal} is nearly optimal, as we discuss in this paper. We have shown that in the general case where all four additional parameters are left free, the joint analysis is not useful for Planck: it does not give better constraints than the power spectrum alone. We also gave a theoretical argument for why this must be so with no detection of isocurvature modes in the power spectrum and no detection of non-Gaussianity (any constraints coming from the bispectrum can be absorbed by the $\kappa$ in this case).\par 

However, if we consider a more restricted class of models where either $\cos\Delta$ or the $\kappa$'s are fixed to a specific non-zero value (certain curvaton models predict for example $\cos\Delta=\pm 1$), then the joint analysis can improve the constraints even in the case of Planck. In particular we showed that for $|\kappa|>10^3$ fixed, the joint analysis will give better constraints on $\beta_{\rm iso}$ and $\cos\Delta$ than the power spectrum alone. The larger $\kappa$ is, the smaller the allowed interval of those parameters around zero is, and hence the closer to a pure $\Lambda$CDM cosmology we are. Similarly, for a fixed value of $|\cos\Delta|\leq 0.1$ (but distinct from zero) the joint analysis improves the constraints on $\beta_{\rm iso}$ (and pushes the most likely value of $\kappa$ upwards). Remarkably, for such values of $\cos\Delta$ in the case of the neutrino velocity isocurvature mode, the joint analysis even seems to indicate a detection of $\beta_{\rm iso}$ at the level of $3$--$4\sigma$. However, because of different reasons including the differences between the Fisher error bars and the simulation-based error bars for exactly those $\tilde f_\mathrm{NL}$ components on which this conclusion is based, we consider this to be a statistical fluke.\par

Going beyond Planck, to future experiments like LiteBIRD and CMB-S4, we use a simplified model of the observations without foreground residuals and a simplified power spectrum likelihood. We should keep in mind that foregrounds could have an impact on parameter estimation by correlating modes in the power spectrum/bispectrum estimation. Although, in the case of Planck, comparisons between forecasts \cite{Langlois:2012tm} and real results have shown that the effect of foreground residuals is small for adiabatic and isocurvature $\tilde f_{\rm NL}$ estimation. Forecasts for isocurvature parameter estimation from the power spectrum, like in \cite{Enqvist:1999vt,Hamann:2009yf,Baumann:2008aq}, are also in good agreement with the real Planck results. However, the impact of foreground residuals in the case of LiteBIRD and CMB-S4, and for the joint analysis, must still be studied carefully in the future, as well as other effects like anisotropic noise, although the scanning strategy of LiteBIRD with a large precession angle will lead to a more uniform coverage than for Planck. For completeness' sake, let us recall here all the other assumptions we made (not only for LiteBIRD and CMB-S4, but also for Planck): all spectral indices are equal, only one isocurvature mode is considered at a time in addition to the adiabatic mode and the correlation of the two, the cosmological parameters are fixed in the bispectrum analysis to find the $\tilde f_{\rm NL}^0$, we assumed a two-field model where only one field contributes to the isocurvature mode and to the non-Gaussianity, as well as statistical independence of the two- and three-point statistics.\par

Our theoretical assessment showed that in the general case (leaving all four parameters free) the joint analysis can improve the constraints if two conditions are satisfied. Firstly, the isocurvature mode amplitude $\beta_{\rm iso}$ must be detected in the power spectrum, otherwise the parameter space to sample is infinite and strongly degenerate, which gives results that are difficult to interpret. Secondly, one of the two $\kappa_I$ must be detected. This means that one must detect either one of the two $\tilde f_{\rm NL}^{I,SS}$ (for $I=\zeta$ or $S$), or both $\tilde f_{\rm NL}^{I,\zeta \zeta}$ and $\tilde f_{\rm NL}^{I,\zeta S}$ with the same first index $I$. If in addition we have a detection of the correlation $\cos\Delta$ in the power spectrum, then even detecting any single $\tilde f_{\rm NL}^{I,JK}$ suffices. We constructed a combined power spectrum and $\tilde f_\mathrm{NL}$ likelihood for LiteBIRD and CMB-S4 and investigated in what region of the $(\beta_{\rm iso}, \cos\Delta)$ parameter space compatible with the Planck results these conditions are satisfied, given also fiducial values for the $\kappa$ parameters compatible with Planck within $1\sigma$. In all our results we found that LiteBIRD is the main driver of the improvements compared to Planck, with CMB-S4 providing only a marginal further improvement.\par

For the CDM isocurvature mode we found that, given the current Planck constraints, the probability of a detection by LiteBIRD+CMB-S4 is unfortunately rather low. We had to choose an unlikely couple of $\beta_{\rm iso}$ and $\cos\Delta$ fiducial values, which are compatible only at $2\sigma$ with Planck. In that case, however, the joint analysis improves the constraints on both $\beta_{\rm iso}$ and $\cos\Delta$ very significantly.\par

For the neutrino isocurvature modes the situation is more hopeful. We can easily find fiducial values for $\beta_{\rm iso}$ and $\cos\Delta$ within the Planck $1\sigma$ contours where the above conditions are satisfied. For the neutrino velocity mode about half of the region within the Planck $1\sigma$ contour even satisfies these conditions. Our chosen fiducial values mean that $\beta_{\rm iso}$ would be detected by LiteBIRD+CMB-S4 in the power spectrum with $5\sigma$ and $7\sigma$ for neutrino density and velocity, respectively. The joint analysis will then provide very significant improvements on the error bars of $\cos\Delta$ compared to the power spectrum alone. To give an example for the neutrino velocity isocurvature mode, for our chosen fiducial values the error bar of $\cos\Delta$ improves by $67\%$, leading to a highly significant detection at $12\sigma$.\par

We have shown that in particular $\cos\Delta$ is correlated with the standard cosmological parameters $A_s$, $n_s$, and $\theta_{MC}$. Hence, the improvement of its error bars with the joint analysis as discussed above, can induce a non-negligible improvement in these parameters. For the configuration we studied we find for example improvements of the error bars of $A_s$ and $\theta_{MC}$ of about $20\%$ and about $30\%$ for neutrino density and neutrino velocity, respectively, compared to an analysis of the power spectrum alone of the $\Lambda$CDM + one isocurvature mode cosmology.\par

While the main focus of this paper was to investigate improvements of the isocurvature constraints using a joint analysis of the power spectrum and the bispectrum, we also discussed three simple examples of possible future measurements by LiteBIRD+CMB-S4 that would rule out the general class of two-field inflation models that we assumed for the joint analysis. The model predicts certain relations between the $\tilde f_{\rm NL}$ parameters themselves (in particular regarding their signs), as well as between the $\tilde f_{\rm NL}$ and the power spectrum parameters $\beta_\mathrm{iso}$ and $\cos\Delta$. As we showed, it is possible for the LiteBIRD+CMB-S4 measurements to be incompatible at high confidence level with those relations, while staying within the Planck-allowed region.\par

Of course it is possible that no isocurvature modes will be detected by LiteBIRD+CMB-S4 (and that is even very likely for the CDM isocurvature mode), in which case the joint analysis will be useless for the general four-parameter model. One should also not forget the various assumptions we made in our analyses.
Still, it is interesting to see that for the neutrino isocurvature modes, and in particular for the neutrino velocity mode, there are significant regions of the parameter space compatible with Planck where a detection by LiteBIRD+CMB-S4 is possible, and where the joint analysis can provide a very significant improvement compared to an analysis of the power spectrum alone. In addition, we saw for Planck that in the case of a more restricted model with fewer free parameters, the joint analysis could be useful for improving the constraints even without a detection. While we will leave forecasts for LiteBIRD+CMB-S4 for those more restricted models to future work, it seems reasonable to expect similar results in that case.

\acknowledgments We thank Josquin Errard for help and explanations regarding LiteBIRD and CMB-S4, as well as the anonymous referee for useful comments. We gratefully acknowledge the IN2P3 Computer Center
(\url{https://cc.in2p3.fr}) for providing the computing resources and services needed for the analysis. TM acknowledges the support of the Paris Centre for Cosmological Physics.

\bibliographystyle{JHEP.bst}

\bibliography{references.bib}

\end{document}